\documentclass[11pt,a4paper]{article}
\usepackage{amsmath,amsfonts,epsfig}
\usepackage{amssymb}
\usepackage{mathrsfs}
\usepackage{hyperref}
\usepackage{amsxtra}
\usepackage{amscd}
\usepackage{amsthm}
\usepackage{amsfonts}
\usepackage{amssymb}
\usepackage{eucal}
\usepackage{graphicx}
\usepackage{graphics,rotating}
\begin{document}
\numberwithin{equation}{section}
\setlength{\unitlength}{.8mm}

\begin{titlepage} 
\vspace*{0.5cm}
\begin{center}
{\Large\bf Finite volume expectation values in the sine-Gordon model}
\end{center}
\vspace{1.5cm}
\begin{center}
{\large \'Arp\'ad Heged\H us}
\end{center}
\bigskip

\vspace{0.1cm}

\begin{center}
Wigner Research Centre for Physics,\\
H-1525 Budapest 114, P.O.B. 49, Hungary\\ 
\end{center}
\vspace{1.5cm}
\begin{abstract}
Using the fermionic basis discovered in the 6-vertex model, we derive exact formulas for the expectation values 
of local operators of the sine-Gordon theory in any eigenstate of the Hamiltonian. 
We tested our formulas in the pure multi-soliton sector of the theory.   
In the ultraviolet limit, we checked our results against  Liouville  3-point functions, 
while in the infrared limit, we evaluated our formulas in the semi-classical limit and compared them upto 2-particle contributions 
against the semi-classical limit of the previously conjectured LeClair-Mussardo type formula. 
Complete agreement was found in both cases.
\end{abstract}

\end{titlepage}

\section{Introduction} \label{intro}

The knowledge of finite volume form-factors of integrable quantum field theories 
became important in string theory and in condensed matter applications, as well.
In string-theory, they arise in the AdS/CFT correspondence when heavy-heavy-light 
3-point functions are considered \cite{BJsftv,BJhhl,Hollo,JF1,JF2}. In condensed matter physics the finite volume 
form-factors are necessary to represent correlation functions for describing various quasi 1-dimensional 
condensed matter systems \cite{KoEss}.

So far two basic approaches have been developed to compute finite volume matrix elements of local operators 
in an integrable quantum field theory. In the first approach \cite{PT08a,PT08b}, the finite-volume form-factors are represented 
as a large volume series in terms of the infinite volume form-factors of the theory. In this approach the polynomial 
in volume corrections are given by the Bethe-Yang quantizations of the rapidities, while the exponentially small in volume 
L\"uscher corrections come from contributions of virtual particles propagating around the compact dimension \cite{Pmu,BBCL,BBCL1}.
In a diagonally scattering theory, the diagonal matrix elements of local operators can be computed by 
means of the LeClair-Mussardo series \cite{LM99,Pozsg13,PST14}, provided the 
connected diagonal form-factors of the operator under consideration are known.
In a non-diagonally scattering theory, so far the availability of such a series representation is restricted to the sine-Gordon model \cite{En,En1}.

The second approach, which works well for non-diagonally scattering theories, as well, 
is based on some integrable lattice regularization of the model under consideration.
The so far elaborated models are the sine-Gordon \cite{SGV,En} and the $N=1$ super sine-Gordon theories \cite{SmirBab}. 
The results of \cite{SGV} for the sine-Gordon model made it possible to conjecture exact formulas for the finite volume diagonal 
matrix elements of the sinh-Gordon model \cite{SmirNeg,SmirBajn}, as well.
The main advantage of the lattice approach is that, it provides exact formulas for the (specific ratios of) finite-volume form-factors. 
From this approach, the corresponding LeClair-Mussardo series can be obtained by the large volume series expansion of the formulas. 
In \cite{SGV} the finite volume ground state expectation values of the exponential fields and their descendants have been determined in the sine-Gordon model. 
In this paper we extend these results to get formulas for the expectation values of local operators in any excited state of the theory. 

The outline of the paper is as follows: In section \ref{sect2}, we formulate 
the sine-Gordon model as perturbed 
conformal field theory. In section \ref{sect3}, we review the equations governing the finite volume spectrum of the theory. 
In section \ref{sect40}, we derive equations for the function $\omega,$ which is the fundamental 
building block of the expectation value formulas of  local operators. 
In section \ref{sect4}, we recall from \cite{SGV}, how the expectation values of the exponential fields and their 
descendants are built up from the function $\omega.$
In section \ref{sect5}, we perform the large volume checks of our formulas for the expectation values. 
The most important of them is a  comparison  
 to the classical limit of the previously conjectured \cite{En} LeClair-Mussardo type large volume series representation. 
Section \ref{sect6}, contains the ultraviolet tests of our formulas. Finally, the paper is closed by our conclusions.

\section{Sine-Gordon model as a perturbed conformal field theory} \label{sect2}

In this section, we recall the perturbed conformal field theory (PCFT) descriptions of the sine-Gordon model defined by the Euclidean action: 
\begin{equation} \label{LSG}
\begin{split}
{\cal A}_{SG}=\int \bigg\{ \frac{1}{4 \pi} \partial_z \varphi(z,\bar{z}) \, \partial_{\bar{z}} \varphi(z,\bar{z})
+\frac{2 {\bf \mu}^2}{\sin \pi \beta^2} \cos(\beta \, \varphi(z,\bar{z})) \bigg\} \, \frac{i \, dz \wedge \, d\bar{z}}{2},
\end{split}
\end{equation}
where $z=x+i\, y$ and $\bar{z}=x-i \, y,$ with $x,y$ being the coordinates of the Euclidean space-time.

In the paper, we will use two perturbed conformal field theory formulations of this model. 
The first one is, when it is considered as a perturbed complex Liouville CFT, and the second one 
is when it is described as a perturbed $c=1$ compactified boson.

In the original paper \cite{SGV}, the formulation with complex Liouville CFT is used. Nevertheless, it turned out from the  detailed \cite{FRT1,FRT2,FRT3}
UV analysis of the finite volume spectrum, that all eigenstates of the Hamiltonian are in one to one correspondence to the operators of 
the $c=1$ modular invariant compactified boson CFT. Thus, we find eligible to describe the finite volume form factors of the operators 
corresponding to the $c=1$ PCFT description of the model.

In the later sections, we will see that in some sense, this set of operators plays a special role among the operators of the complex Liouville-theory. 
Namely, in the exact description of finite-volume form-factors \cite{SGV}, only the ratios of diagonal  form-factors can be computed.  
But, for the operators corresponding to the $c=1$ CFT, the explicit expectation values themselves can be computed exactly, and not only their ratios.


\subsection{Perturbed Liouville CFT formulation}\label{sect2a}

In \cite{SGV}, the sine-Gordon model is considered as a perturbed complex Liouville theory:
\begin{equation} \label{PCLT}
\begin{split}
{\cal A}_{SG}={\cal A}_{L}+ \frac{{\bf \mu}^2}{\sin \pi \beta^2} \int e^{-i \, \beta \, \varphi(z,\bar{z})} \frac{i \, dz \wedge \, d\bar{z}}{2},\quad 
\end{split}
\end{equation}
where ${\cal A}_L$ denotes the action of the complex Liouville CFT:
\begin{equation} \label{AL}
\begin{split}
{\cal A}_{L}=\int \bigg\{ \frac{1}{4 \pi} \partial_z \varphi(z,\bar{z}) \, \partial_{\bar{z}} \varphi(z,\bar{z})
+\frac{ {\bf \mu}^2}{\sin \pi \beta^2} e^{i \, \beta \, \varphi(z,\bar{z})} \bigg\} \, \frac{i \, dz \wedge \, d\bar{z}}{2}.
\end{split}
\end{equation}
The central charge of the CFT is
\begin{equation} \label{centLiou}
\begin{split}
c_L=1-6 \frac{\nu^2}{1-\nu}, \qquad \nu=1-\beta^2,
\end{split}
\end{equation}
where we introduced the parameter $\nu=1-\beta^2$ in order to fit to the notations of ref. \cite{SGV}.
We just mention, that $0<\nu<1$ is the range of the parameter, such that the ranges $\tfrac{1}{2}<\nu<1,$ and 
$0<\nu<\frac{1}{2}$ correspond to the attractive and repulsive regimes of the model. 
The primary fields are labeled by the real continuous parameter $\alpha$:
\begin{equation} \label{primLiou}
\begin{split}
\Phi_\alpha(z,\bar{z})=e^{\tfrac{i\, \alpha \beta \nu}{2 (1-\nu)}  \, \varphi(z,\bar{z})},
\end{split}
\end{equation}
and have scaling dimensions $2 \Delta_\alpha$ with:
\begin{equation} \label{Dalfa}
\begin{split}
\Delta_\alpha=\frac{\nu^2}{4(1-\nu)} \alpha \, (\alpha-2).
\end{split}
\end{equation}
Primary fields (\ref{primLiou}) and their descendants span the basis in the 
space of operators of the theory.

\subsection{Compactified free boson description}\label{sect2b}

The sine-Gordon theory can be formulated as the perturbation of a free compactified boson CFT. Now, the whole potential term of  
 (\ref{LSG}) plays the role of the perturbation:
\begin{equation} \label{PBform}
\begin{split}
{\cal A}_{SG}={\cal A}_{B}+ \frac{ 2 {\bf \mu}^2}{\sin \pi \beta^2} \int \cos(\beta \, \varphi(z,\bar{z})) \frac{i \, dz \wedge \, d\bar{z}}{2},\quad  
\end{split}
\end{equation}
where ${\cal A}_B$ denotes the action of the free boson compactified on a circle of radius $R=\tfrac{1}{\beta}:$
\begin{equation} \label{Abe}
\begin{split}
{\cal A}_{B}=\int  \frac{1}{4 \pi} \partial_z \varphi(z,\bar{z}) \, \partial_{\bar{z}} \varphi(z,\bar{z})
 \, \frac{i \, dz \wedge \, d\bar{z}}{2}.
\end{split}
\end{equation}
The primary states of this CFT are created by the vertex operators $V_{n,m}(z,\bar{z})$ which  
are labeled by two quantum numbers $n \in {\mathbb R}$ and $m \in {\mathbb Z}.$ Their 
left and right conformal dimensions are given by:
\begin{equation} \label{Deltapm}
\begin{split}
\Delta^{\pm}_{n,m}=\left(\frac{n}{R} \pm \frac{1}{4} m R \right)^2.
\end{split}
\end{equation}
Here $n$ is the momentum quantum number, and $m$ is the winding number or topological charge.
The requirement of the locality of the operator product algebra of the CFT imposes further severe 
restrictions on the possible values of the pair of quantum numbers $(n,m).$ It turns out \cite{klassme}, that only a 
bosonic and a fermionic  maximal subalgebras of the vertex operators $V_{n,m}(z,\bar{z})$ are allowed.  
The  bosonic subalgebra is characterized by the quantum numbers $\{n \in {\mathbb Z}, m \in {\mathbb Z}\}.$  
It corresponds to the modular invariant partition function and this CFT describes the UV limit of the sine-Gordon model \cite{klassme}.

In the fermionic subalgebra the allowed set of quantum numbers is given by 
$\{n \in {\mathbb Z}, m \in 2{\mathbb Z}\} \cup \{ n \in {\mathbb Z}+\tfrac{1}{2}, m \in 2{\mathbb Z}+1\}.$ 
It corresponds to a $\Gamma_2$ invariant partition function and this CFT describes the UV limit of the massive-Thirring model \cite{klassme}. 

The perturbing term in the action (\ref{PBform}) is given in terms of these vertex operators as follows:
\begin{equation} \label{Apert}
\begin{split}
{\cal A}_{pert}=\frac{\mu^2}{\sin \pi \beta^2}\int \, \frac{i \, dz \wedge \, d\bar{z}}{2}  \left( V_{1,0}(z,\bar{z})+V_{-1,0}(z,\bar{z})\right).
\end{split}
\end{equation}

In this paper we will mostly focus on computing the diagonal matrix elements of the primaries and their descendants belonging to the bosonic subalgebra of the $c=1$ CFT.

Diagonal matrix elements are non-zero only in the zero winding number sector $(m=0),$ thus our primary goal is to derive formulas for the diagonal matrix elements of the 
vertex operators $V_{n,0}(z,\bar{z}),$ and of their descendants with $n \in {\mathbb Z}.$
In the Liouville formulation they correspond to the primaries $\Phi_{2 \, n \, \tfrac{1-\nu}{\nu}}(z,\bar{z})$ and their descendants.

\section{Integral equations for the spectrum} \label{sect3}

In this section, we summarize the nonlinear integral equations governing the finite volume spectrum of the sine-Gordon theory. 
The equations are derived from an inhomogeneous 6-vertex model, which serves as an 
integrable light-cone lattice regularization \cite{ddvlc} of the sine-Gordon model. 
The derivation of the equations can be found in the papers 
\cite{KP1}-\cite{FRT3}.{\footnote{For the final, in every respect correct form of the equations see  \cite{FevPhd}.}}  
The unknown-function of the nonlinear integral equations (NLIE) is the counting-function of the 6-vertex model{\footnote{To be more precise, 
what we call counting-function in this paper, is the continuum-limit of the counting-function of the 6-vertex model.}}.  
The counting-function $Z(\lambda)$ is an $i \pi$ periodic function. 
In the fundamental regime: $|\text{Im}(\lambda)|<\text{min}(\pi \nu,\pi(1-\nu)),$ it satisfies the equations  as follows \cite{FRT3};
\begin{equation} \label{ddv1}
\begin{split}
&Z(\lambda)=\ell \, \sinh\tfrac{\lambda}{\nu}+g(\lambda|\{h_j\},\{\lambda_j\},\{y_j\})-\\
&\int\limits_{-\infty}^{\infty} \frac{d\lambda'}{i} \,
R(\lambda-\lambda'-i0|0) \, L_+(\lambda'+i0)+
\int\limits_{-\infty}^{\infty} \frac{d\lambda'}{i} \,
R(\lambda-\lambda'+i0|0) \, L_-(\lambda'-i0),
\end{split}
\end{equation}
where $\ell={\cal M} L$ with ${\cal M}$ and $L$ being the soliton-mass and the size of the compactified dimension, 
\begin{equation} \label{Lpm}
\begin{split}
L_\pm(\lambda)=\ln\left( 1+(-1)^\delta e^{\pm i Z(\lambda)} \right), \qquad \delta\in\{0,1\},
\end{split}
\end{equation}
\begin{equation} \label{ddvg1}
\begin{split}
g(\lambda|\{h_j\},\{\lambda_j\},\{y_j\})=&-2\pi \sum\limits_{j=1}^{m_H} \chi_R(\lambda-h_j)+2\pi \sum\limits_{j=1}^{m_C} \chi_R(\lambda-c_j)+
2\pi \sum\limits_{j=1}^{m_W} \chi_{R,II}(\lambda-w_j)+\\
&2\pi \sum\limits_{j=1}^{m_S} (\chi_R(\lambda-y_j-i \, 0)+\chi_R(\lambda-y_j+i \, 0)).
\end{split}
\end{equation}
The functions $R$ and $\chi_R$ entering the equations are of the form:
\begin{equation} \label{Rfv}
\begin{split}
R(\lambda|\alpha)=\int\limits_{-\infty}^{\infty} \frac{d \omega}{2 \pi} e^{i \omega \lambda} \tilde{R}(\omega|\alpha), \qquad 
\tilde{R}(\omega|\alpha)=\frac{1}{2}\frac{\sinh\left[\tfrac{\pi \omega}{2}(2 \nu-1)-\tfrac{i \pi \alpha}{2}\right]}{\cosh\tfrac{\pi \nu \omega}{2} \,\sinh\left[\tfrac{\pi \omega}{2}( 1-\nu)+\tfrac{i \pi \alpha}{2}\right] },
\end{split}
\end{equation}
\begin{equation} \label{chiR}
\begin{split}
\chi_R(\lambda)=\int\limits_{-\infty}^{\infty} \frac{d \omega}{2 \pi} \frac{\sinh(\omega \, \lambda)}{\omega} \tilde{R}(\omega|0).
\end{split}
\end{equation}
The label $II$ on any function means second determination with the definition as follows \cite{ddv97}:
\begin{equation} \label{fII}
\begin{split}
f_{II}(\lambda)=\left\{ 
\begin{array}{r}
f( \lambda)+f(\lambda-i \, \pi \,\nu \text{sign}(\text{Im} \lambda)), \qquad \qquad 
\qquad 0<\nu<\tfrac{1}{2}, \\
f( \lambda)-f(\lambda-i \, \pi \,(1-\nu) \, \text{sign}(\text{Im} \lambda)).
\qquad \qquad  \qquad \tfrac{1}{2}<\nu<1.
\end{array}\right.
\end{split}
\end{equation}
The objects $\{h_j\},\{\lambda_j\},\{y_j\}$ entering the source term (\ref{ddvg1}), are  
zeros of the nonlinear expression $1+(-1)^\delta \, e^{i Z(\lambda)},$ 
with the properties as follows:
\begin{itemize}
\item holes: $h_k \in \mathbb{R}, \quad k=1,...,m_H$ are those real solutions of $1+(-1)^\delta \, e^{i Z(h_j)}=0,$ which are not Bethe-roots.
\item close roots: $c_k, \quad k=1,...,m_C$, with $0<|\mbox{Im} c_k|\leq \mbox{min}(\pi \nu, \pi (1-\nu))$, are complex solutions of 
$1+(-1)^\delta \, e^{i Z(c_j)}=0.$ They are complex zeros of the Bethe-equations. 
\item wide roots: $w_k, \quad k=1,...,m_W$, with $\mbox{min}(\pi \nu,\pi (1-\nu))<|\mbox{Im} w_k|\leq \tfrac{\pi}{2}$, are complex Bethe-roots 
satisfying the equation 
$1+(-1)^\delta \, e^{i Z(w_j)}=0.$ 
\item special objects
: $y_k \in \mathbb{R}, \quad k=1,...,m_S \,\,$ are either holes or real Bethe-roots 
satisfying the equation  $1\!+\!(-1)^{\delta}e^{i Z(y_k)}\!=\!0$ with $Z'(y_k)<0.$
\end{itemize}
The topological charge $Q$ of the state described by these objects, is given by the so-called counting-equation \cite{FRT3}:
\begin{equation} \label{counteq}
\begin{split}
Q=m_H-2 m_S-m_C-2 H(1/2-\nu)\, m_W,
\end{split}
\end{equation}
where here $H(x)$ denotes the Heaviside-function.
The equations satisfied by the source objects can be rephrased in their logarithmic form, as well:
\begin{eqnarray}
&\bullet& \, \text{holes:}  \qquad Z(h_k)=2\pi \, I_{h_k}, \qquad I_{h_k} \in \mathbb{Z}+\tfrac{1+\delta}{2}, \qquad k=1,..,m_H, \label{Hkvant}\\
&\bullet& \, \text{close roots:}  \qquad Z(c_k)=2\pi \, I_{c_k}, \qquad I_{c_k} \in \mathbb{Z}+\tfrac{1+\delta}{2}, \qquad k=1,..,m_C, \label{Ckvant} \\
&\bullet& \, \text{wide roots:}  \qquad Z(w_k)=2\pi \, I_{w_k}, \qquad I_{w_k} \in \mathbb{Z}+\tfrac{1+\delta}{2}, \qquad k=1,..,m_W, \label{Wkvant} \\
&\bullet& \, \text{special objects:}  \qquad Z(y_k)=2\pi \, I_{y_k}, \qquad I_{y_k} \in \mathbb{Z}+\tfrac{1+\delta}{2}, \qquad k=1,..,m_S. \label{Skvant}
\end{eqnarray}
This formulation shows, that the actual value of the parameter $\delta \in \{0,1\}$ determines whether the source objects are quantized by integer or half integer quantum numbers.  
The choice of $\delta$ is not arbitrary, but should follow the selection rules \cite{FRT3}: 
 \begin{eqnarray} 
&\bullet& \quad \frac{Q+\delta+M_{sc}}{2} \in \mathbb{Z}, \qquad \text{sine-Gordon,} \qquad \qquad\qquad \qquad \qquad\qquad \label{SGkvant} \\
&\bullet& \quad \frac{\delta+M_{sc}}{2} \in \mathbb{Z}, \qquad  \text{massive Thirring,} \qquad \qquad\qquad \qquad \qquad\qquad \label{MTkvant},
\end{eqnarray}
where here $M_{sc}$ stands for the number of self-conjugate roots, which are such wide roots, whose imaginary parts are fixed by the periodicity of $Z(\lambda)$ to 
$ \tfrac{\pi}{2}.$ Here we recall, that due to the periodicity of $Z(\lambda),$ the complex roots can be restricted to the domain: 
$-\tfrac{\pi}{2}<\lambda_j \leq \tfrac{\pi}{2}.$
The form (\ref{ddv1}) is appropriate to impose the quantization equations (\ref{Hkvant}), (\ref{Ckvant}),(\ref{Skvant}), but to impose 
the quantization equations (\ref{Wkvant}) for the wide roots, the form of (\ref{ddv1}) should be analytically continued to the strip 
$\text{min}(\pi \nu,\pi\, (1-\nu))<\text{Im} \, \lambda\leq \tfrac{\pi}{2},$ as well. 
In this domain the equations take the form as follows \cite{ddv97,FRT1,FRT2,FRT3}:
\begin{equation} \label{ddv2}
\begin{split}
&Z(\lambda)=\ell \, \sinh_{II}\tfrac{\lambda}{\nu}+g(\lambda|\{h_j\},\{\lambda_j\},\{y_j\})_{II}-\\
&\int\limits_{-\infty}^{\infty} \frac{d\lambda'}{i} \,
R_{II}(\lambda-\lambda'-i0|0) \, L_+(\lambda'+i0)+
\int\limits_{-\infty}^{\infty} \frac{d\lambda'}{i} \,
R_{II}(\lambda-\lambda'+i0|0) \, L_-(\lambda'-i0),
\end{split}
\end{equation}
where $II$ denotes the second determination (\ref{fII}).

The energy and momentum of the eigenstates in units of the soliton mass are given by the expressions as follows:  
\begin{equation} \label{EL}
\begin{split}
\frac{E(L)\!-\!E_{b}\,\ell}{\cal M}&=\sum\limits_{j=1}^{m_H} \cosh \tfrac{h_j}{\nu}-\sum\limits_{j=1}^{m_C} \cosh \tfrac{c_j}{\nu}-
\sum\limits_{j=1}^{m_W} \cosh_{II} \tfrac{w_j}{\nu}
-\sum\limits_{j=1}^{m_S} \left\{ \cosh \tfrac{y_j+i 0}{\nu}+\cosh \tfrac{y_j-i 0}{\nu} \right\}\\
&-\int\limits_{-\infty}^{\infty} \!\! \frac{d \lambda}{2 \pi \nu i} \, \sinh \tfrac{\lambda+i 0}{\nu} \, L_+(\lambda+i 0)
+\int\limits_{-\infty}^{\infty} \!\! \frac{d \lambda}{2 \pi \nu i} \, \sinh \tfrac{\lambda-i 0}{\nu} \, L_-(\lambda-i 0),
\end{split}
\end{equation}
\begin{equation} \label{PL}
\begin{split}
\frac{P(L)}{\cal M}&=\sum\limits_{j=1}^{m_H} \sinh \tfrac{h_j}{\nu}-\sum\limits_{j=1}^{m_C} \sinh \tfrac{c_j}{\nu}-
\sum\limits_{j=1}^{m_W} \sinh_{II} \tfrac{w_j}{\nu}
-\sum\limits_{j=1}^{m_S} \left\{ \sinh \tfrac{y_j+i 0}{\nu}+\sinh \tfrac{y_j-i 0}{\nu} \right\}\\
&-\int\limits_{-\infty}^{\infty} \!\! \frac{d \lambda}{2 \pi \nu i} \, \cosh \tfrac{\lambda+i 0}{\nu} \, L_+(\lambda+i 0)
+\int\limits_{-\infty}^{\infty} \!\! \frac{d \lambda}{2 \pi \nu i} \, \cosh \tfrac{\lambda-i 0}{\nu} \, L_-(\lambda-i 0),
\end{split}
\end{equation}
where $L_\pm(\lambda)$ is defined in (\ref{Lpm}) and $E_{b}=\tfrac{\cal M}{4}\,\cot \tfrac{\pi}{2 \nu}$ is the 
bulk energy constant.

\section{The function $\omega$ for excited states} \label{sect40}

As it will be summarized in the next section, the expectation values of all primaries (\ref{primLiou}) and their 
Virasoro descendants can be expressed in terms of a single function $\omega(\mu,\lambda|\alpha)$ \cite{SGV}. 
This function depends on two spectral parameters $\mu,\lambda,$ a twist parameter $\alpha,$ and although 
it is not indicated, it depends on the state which sandwiches the operators. Thus it can be considered as a functional 
of the counting-function of the sandwiching state. The derivation of this function for the ground state can be found in 
papers \cite{CFT4,SGV}. The result for the sine-Gordon model takes the relatively simple form \cite{SGV}:
 \begin{equation} \label{omgs}
\begin{split}
\frac{\omega(\mu,\lambda|\alpha)}{4\pi i}=-(F^+\star F^-)(\mu,\lambda)+(F^+\star R_d \star F^-)(\mu,\lambda),
\end{split}
\end{equation}
where
\begin{equation} \label{F+-}
\begin{split}
F^\pm(\lambda)=\frac{i}{4 \pi \nu} \frac{1}{\sinh \tfrac{\lambda\mp i\, 0}{\nu}},
\end{split}
\end{equation}
$R_d(\mu,\lambda)$ is the solution of the equation:
\begin{equation} \label{Rdeq}
\begin{split}
R_d(\mu,\lambda)+(R_d \star R)(\mu,\lambda)=R(\mu-\lambda|\alpha),
\end{split}
\end{equation}
where $R(\lambda|\alpha)$ is given in (\ref{Rfv}) and the definition of the $\star$ convolution is as follows:
\begin{equation} \label{star0}
\begin{split}
(f \star g)(\mu,\lambda)=\int\limits_{-\infty}^{\infty} dm(t) \, f(\mu,t)\, g(t,\lambda),
\end{split}
\end{equation}
where the integral measure contains the counting-function:
\begin{equation} \label{mertek}
\begin{split}
&dm(t)=\left( \frac{1}{1+a(t-i \, 0)}+\frac{1}{1+\bar{a}(t+i \, 0)}\right), \\
\qquad &a(\lambda)=(-1)^{\delta} e^{i \, Z(\lambda)}, \qquad 
\bar{a}(\lambda)=\tfrac{1}{a(\lambda)}=(-1)^{\delta} e^{-i \, Z(\lambda)}.
\end{split}
\end{equation}

Equations (\ref{omgs})-(\ref{mertek}) give the function $\omega$ for the ground state of the model. Now, we extend it to 
any excited state of the model. To do so, first we introduce the usual convolution as well:
\begin{equation} \label{conv}
\begin{split}
(f * g)(\mu,\lambda)=\int\limits_{-\infty}^{\infty} \! dt \, f(\mu,t+i \, 0)\, g(t+i \, 0,\lambda).
\end{split}
\end{equation}
If one starts to analyze the singularity structure of the function $\tfrac{1}{1+a(\lambda)},$ it turns out, 
that it has simple poles at the positions of holes or at the 
positions of the Bethe-roots with residues $\tfrac{1}{a'(\lambda)}$ taken at the position of the singularity.
It is known, that the ground state is a state without any hole,
such that all Bethe-roots are real. Thus the $\star$ convolution can be written in the following equivalent way:
\begin{equation} \label{star1}
\begin{split}
(f \star g)(\mu,\lambda)=\oint\limits_{\gamma} \, \frac{1}{1+a(t)} \, f(\mu,t)\, g(t,\lambda)-(f * g)(\mu,\lambda),
\end{split}
\end{equation}
where $\gamma$ is a contour encircling all the Bethe-roots, but none of the holes. But this formulation of the convolution can be easily extended to excited states, 
 only the contributions of holes and of the complex Bethe-roots should be taken into account by the appropriate application of the residue theorem, when the contours 
 are deformed to run parallel to the real axis. Thus for a generic excited state, the $\star$ convolution takes the form as follows: 
\begin{equation} \label{star2}
\begin{split}
(f \star g)(\mu,\lambda)\!=\!\!\!\int\limits_{-\infty}^{\infty} \! dm(t) \, f(\mu,t)\, g(t,\lambda)\!-\!2 \pi  i \sum\limits_{j=1}^{m_H} \! \frac{f(\mu,h_j) \, g(h_j,\lambda)}{a'(h_j)}
\!+\!
2 \pi i \sum\limits_{j=1}^{m_K} \! \frac{f(\mu,\lambda_j) \, g(\lambda_j,\lambda)}{a'(\lambda_j)},
\end{split}
\end{equation}
where $\lambda_j$ denotes the complex Bethe-roots, which are exactly the complex Bethe-roots of the NLIE (\ref{ddv1}): 
$\{\lambda_j\}_{j=1}^{m_K}=\{c_j\}_{j=1}^{m_C} \cup \{w_j\}_{j=1}^{m_W}.$ 
The only point where the counting-function, which characterizes the sandwiching state, arises 
in the formulas for $\omega,$ is the $\star$ convolution. 
Consequently, the original formulas of (\ref{omgs})-(\ref{Rdeq}) of \cite{SGV} remain unchanged for excited states, provided the new 
definition (\ref{star2}) is used for the $\star$ convolution. The only subtle point, which should be clarified is how to interpret 
the function $R(\mu,\lambda|\alpha)$ in (\ref{Rdeq}), when its arguments have large enough imaginary parts. To clarify this point,  
one needs to use another representation of (\ref{Rdeq}), which defined $R_d(\mu, \lambda)$ 
for arbitrary complex values of its arguments{\footnote{See (3.9) in \cite{CFT4}.}} \cite{CFT4};
\begin{equation} \label{RdK}
\begin{split}
R_d(\mu,\lambda)+\oint\limits_{\gamma} \, \frac{1}{1+a(t)} \, R_d(\mu,t)\, K_\alpha(t-\lambda)=K_\alpha(\mu-\lambda), \qquad  \mu,\lambda \in {\mathbb C},
\end{split}
\end{equation}
where here the curve $\gamma$ is the same as in (\ref{star1}), the function $K_\alpha(\lambda)$ is given by:
\begin{equation} \label{Kalf}
\begin{split}
K_\alpha(\lambda)=\frac{e^{\alpha \lambda}}{2 \pi i} \, \left(e^{i \pi \nu \alpha} \left[ \coth(\lambda+i \pi \nu)-1\right] 
-e^{-i \pi \nu \alpha} \left[ \coth(\lambda-i \pi \nu)-1\right]\right),
\end{split}
\end{equation}
or equivalently in Fourier space:
\begin{equation} \label{KalfF}
\begin{split}
K_\alpha(\lambda)=\int\limits_{-\infty}^{\infty} \frac{d\omega}{2 \pi} e^{i \omega \lambda} \tilde{K}_\alpha(\omega), 
\qquad \tilde{K}_\alpha(\omega)=\frac{1}{2}\frac{\sinh\left[\tfrac{\pi \omega}{2}(2 \nu-1)-\tfrac{i \pi \alpha}{2}\right]}{\,\sinh\left[\tfrac{\pi \omega}{2}+\tfrac{i \pi \alpha}{2}\right] }.
\end{split}
\end{equation}
Then using (\ref{star1}) and acting on (\ref{RdK}) with $(1-K_\alpha)^{-1}$ from the right by means of the usual $*$ convolution, one ends up with (\ref{Rdeq}). 
The only subtlety is that one has to take into account the poles of $K_{\alpha}(\lambda),$ when the arguments $\mu$ and $\nu$ have large enough imaginary parts. 
The detailed study of all possible cases shows, that the form of (\ref{Rdeq}) remains the valid for arbitrary complex values of 
$\lambda$ and $\mu$, provided, the kernel $R(\mu,\nu|\alpha)$ in (\ref{Rdeq}) is defined by the formula:
\begin{equation} \label{Rfvall}
\begin{split}
R(\mu,\lambda|\alpha)\!=\!K_\alpha(\mu-\lambda)\!+\!\!\!\!\int\limits_{-\infty}^{\infty} \!\! d\tau K_\alpha(\mu\!-\!\tau) K_\alpha(\tau\!-\!\lambda)\!
+\!\!\!\! \int\limits_{-\infty}^{\infty} \!\! d\tau \, d\tau' \,
K_{\alpha}(\mu\!-\! \tau) R(\tau\!-\!\tau'|\alpha)  K_\alpha(\tau'\!-\!\lambda),
\end{split}
\end{equation}
where on the right hand side $R(\tau|\alpha)$ is given by the Fourier integral (\ref{Rfv}). 
When $\mu$ and $\lambda$ are close to the real axis, from the identity, that $R=K_{\alpha}*(1-K_\alpha)^{-1}$, 
it can be easily seen, that this definition gives back the original 
definition (\ref{Rfv}) for $R(\lambda|\alpha)$ in the appropriate neighborhood of the real axis.

As it will be clear from the next section, to compute expectation values of local operators, one does not need to have the functional form of $\omega(\mu,\lambda|\alpha),$ 
but only some of the coefficients of its small and/or large argument series representation. 
Thus, it is worth to define the matrix $\omega_{2j-1,2k-1}(\alpha)$ with the definition \cite{SGV}: 
\begin{equation} \label{ommatr}
\begin{split}
\omega(\mu,\lambda|\alpha)\!=\!\!\!\sum\limits_{j,k=1}^{\infty} \!\!e^{-\epsilon_1 \mu \tfrac{2j-1}{\nu}} e^{-\epsilon_2 \lambda \tfrac{2k-1}{\nu}} 
\omega_{\epsilon_1 (2j-1),\epsilon_2(2k-1)}(\alpha),
\quad \! \epsilon_{1,2}\!=\!\pm, \quad \epsilon_1 \, \mu \!\to \!\infty, \quad \!
\epsilon_2 \, \lambda \! \to \! \infty.
\end{split}
\end{equation}
Fortunately, to compute the matrix elements $\omega_{2j-1,2k-1},$ it is not necessary to solve the equation (\ref{Rdeq}) for the two-argument function $R_d.$ 
It is enough to solve linear equations for functions of a single variable. 
Define the function:
\begin{equation} \label{ejfv}
\begin{split}
e_j(\lambda)=\left\{ 
\begin{array}{r}
e^{j\tfrac{ \, \lambda}{\nu}}, \qquad \qquad \qquad \qquad
\qquad |\text{Im} \lambda|<\text{min} [\pi \nu,\pi(1-\nu)], \\
\left(e^{j\tfrac{ \, \lambda}{\nu}}\right)_{II},
\qquad \qquad  \qquad \text{min} [\pi \nu,\pi(1-\nu)]<|\text{Im} \lambda|<\tfrac{\pi}{2}.
\end{array}\right.
\end{split}
\end{equation}
Let ${\cal G}_j(\lambda)$ be the solution of the linear integral equation as follows;
\begin{equation} \label{Gj}
\begin{split}
{\cal G}_j^{(2)}+R\star {\cal G}_j^{(2)}=e_j^{(2)}.
\end{split}
\end{equation}
Here for short we introduced an upper index in order to be able to use the short notation (\ref{star2}) for the convolution. 
Thus the interpretation of (\ref{Gj}) is as follows, each function ${\cal G}_j,R, e_j$ in (\ref{Gj}) can be considered as two-argument functions, but the 
upper index $"(2)"$ means that ${\cal G}_j$ and $e_j$ are constant in their second argument, i.e. they are functions of only one variable.  
Then from (\ref{omgs}) and (\ref{Rdeq}), it can be shown, that the matrix elements of $\omega$ can be expressed in terms of solutions of (\ref{Gj}) as follows:
\begin{equation} \label{omkif}
\begin{split}
\omega_{\epsilon_1 (2j-1),\epsilon_2(2k-1)}(\alpha)=\frac{-i}{2 \pi \nu^2} \, e_{\epsilon_1 (2j-1)}^{(1)} \star {\cal G}_{\epsilon_2 (2k-1)}^{(2)}, \qquad 
\epsilon_{1,2}=\pm, \quad j,k\in{\mathbb Z}.
\end{split}
\end{equation}
The main advantage of this formula is that, it requires to solve integral equations for  
functions with a single argument, which is much simpler, than to determine 
the full function $\omega(\mu,\nu|\alpha)$ from (\ref{omgs}) and (\ref{Rdeq}).

\section{Formulas for the expectation values} \label{sect4}

Now, we are in the position to summarize how to compute expectation values of local operators in the fermionic basis 
in terms of the single function $\omega(\mu,\lambda|\alpha)$ defined in the previous section.

Local operators of the theory are labeled by their conformal counterparts. The fermionic basis spans the space of 
local operators modulo the action of the integrals of motion \cite{SGV}.

In \cite{CFT4,SGV} it was discovered, that there exist an anti-commuting set of operators acing on the space of local operators of the theory, which allows one to 
construct an alternative to the Virasoro basis in the Verma-module. 

There are two types of fermions for each chirality. The creation and annihilation operators are denoted by $\beta^*_j, \gamma_j^*$, and $\beta_j, \gamma_j$ respectively 
for one chirality, and by $\bar{\beta}^*_j, \bar{\gamma}_j^*$, and $\bar{\beta}_j, \bar{\gamma}_j$ for the other.
The fermions satisfy the anti-commutation relations as follows:
\begin{equation} \label{fermant}
\begin{split}
\{\beta_j,\beta^*_k\}=\{\bar{\gamma}_j,\bar{\gamma}_k^*\}=\delta_{jk} \, t_k(\alpha), \qquad t_k(\alpha)=\tfrac{1}{i\, \nu} \, \cot \tfrac{\pi}{2 \nu}(\nu \, \alpha+k).
\end{split}
\end{equation}
They span the basis of the Verma module{\footnote{More precisely the Verma module factored out by the action of the local integrals of motion \cite{SGV}.}} 
created by the primary $\Phi_{\alpha+2m \tfrac{1-\nu}{\nu}}(0),$ as follows \cite{SGV}:
\begin{equation} \label{basis}
\begin{split}
\beta^*_{I^+}\, \bar{\beta}^*_{\bar{I}^+} \, \bar{\gamma}^*_{\bar{I}^-}  \, \gamma^*_{I^-}\,  \Phi_\alpha^{(m)}(0),
\end{split}
\end{equation}
where $\Phi_\alpha^{(m)}(0)$ denotes the m-fold screened primary field  
\cite{DotFat}, 
 $I^\pm$ and $\bar{I}^\pm$ stand for multi-indexes, namely:
\begin{equation} \label{multii}
\begin{split}
I^+=\{2 i_1^+-1,...,2i_p^+-1\},\quad \beta_{I^+}=\beta^*_{2i_1^+-1}...\beta^*_{2i_p^+-1}, \qquad 1\leq i_1^+<...< i_p^+, \\
I^-=\{2 i_1^--1,...,2i_q^--1\},\quad \gamma_{I^-}=\gamma^*_{2i_q^--1}...\gamma^*_{2i_1^--1}, \qquad 1\leq i_1^-<...< i_q^-,
\end{split}
\end{equation}
and similarly for the other chirality. For a multi-index $I,$ let $\#(I)$ the number of quantum numbers in it. Then in (\ref{basis}) the following constraints 
should hold: $\#(I^+)=\#(I^-)+m,$ $\,\#(\bar{I}^-)=\#(\bar{I}^+)+m.$ 
The lowest dimensional element of the set (\ref{basis}) gives the primary field:
\begin{equation} \label{primrep}
\begin{split}
\Phi_{\alpha+2m\tfrac{1-\nu}{\nu}}(0)\simeq C_m(\alpha) \, \beta^*_{I_{odd}(m)} \, \bar{\gamma}^*_{I_{odd}(m)} \, \Phi_\alpha^{(m)}(0), 
\end{split}
\end{equation}
where 
\begin{equation} \label{Iodd}
\begin{split}
I_{odd}(m)=\{1,3,...,2m-1\},
\end{split}
\end{equation}
and $C_m(\alpha)$ is a constant given in \cite{SGV} by the formula for $m>0$:
\begin{equation} \label{Cm}
\begin{split}
C_m(\alpha)=\prod\limits_{j=0}^{m-1} C_1(\alpha+2j\tfrac{1-\nu}{\nu}),
\end{split}
\end{equation}
where
\begin{equation} \label{C1}
\begin{split}
&C_1(\alpha)=i \, \nu \Gamma(\nu)^{4 x(\alpha)}\, \frac{\Gamma(-2 \nu x(\alpha))}{\Gamma(2 \nu x(\alpha))}\, \frac{\Gamma(x(\alpha))}{\Gamma(x(\alpha)+1/2)} \,
\frac{\Gamma(-x(\alpha)+1/2)}{\Gamma(-x(\alpha))} \cot(\pi x(\alpha)), \\
&\text{with} \qquad x(\alpha)=\frac{\alpha}{2}+\frac{1-\nu}{2 \nu}.
\end{split}
\end{equation}
For $m<0,$ $C_m(\alpha)$ can be determined from the equation \cite{SGV}:
\begin{equation} \label{Cmmin}
\begin{split}
C_{-m}(\alpha) \, C_m(\alpha-2m\tfrac{1-\nu}{\nu})=\nu^{2m}\, \prod_{j=1}^m \tan^2\left( \tfrac{\pi}{2}(\alpha-\tfrac{j}{\nu})\right).
\end{split}
\end{equation}
The main result of \cite{SGV} is that the expectation values of the operators (\ref{basis}) can be expressed in terms of the matrix-elements  (\ref{omkif}) as follows:
\begin{equation} \label{expform}
\begin{split}
&\frac{\langle \beta^*_{I^+}\, \bar{\beta}^*_{\bar{I}^+} \,  \bar{\gamma}^*_{\bar{I}^-} \,
\gamma^*_{I^-}\, \Phi_\alpha^{(m)}(0) \rangle }{\langle \Phi_\alpha(0) \rangle}=\\
&\mu^{2m\alpha-2m^2+\tfrac{1}{\nu}(|I^+|+|I^-|+|\bar{I}^+|+|\bar{I}^-|)} \, {\cal D}_R(I^+\cup (-\bar{I}^+)|I^- \cup (-\bar{I}^-)|\alpha),
\end{split}
\end{equation}
where $|I|$ denotes the sum of elements of $I,$ and for the sets $A=\{a_j\}_{j=1}^n,$ $B=\{b_j\}_{j=1}^n,$ ${\cal D}_R(A|B|\alpha)$ is given by the determinant:
\begin{equation} \label{DR}
\begin{split}
&{\cal D}_R(A|B|\alpha)=\text{det} \, \Omega_{jk}(\alpha), \qquad \text{with} \\
&\Omega_{jk}(\alpha)=\omega_{a_j,b_k}(\alpha)+\tfrac{i}{\nu} \, \text{sign}(a_j) \, \delta_{a_j,-b_k} \, \cot\left(\tfrac{\pi}{2 \nu}(a_j+\nu \alpha)\right).
\end{split}
\end{equation}
Here $\mu$ is the coupling constant in the action (\ref{LSG}) of the sine-Gordon model. It is related to the soliton mass ${\cal M},$ by the formula \cite{ZamiM}:
\begin{equation} \label{muM}
\begin{split}
\mu={\cal M}^\nu \, \Pi(\nu)^\nu, \qquad \text{where} \qquad 
\Pi(\nu)=\frac{\sqrt{\pi}}{2} \frac{\Gamma\left(\tfrac{1}{2 \nu}\right)}{\Gamma\left(\tfrac{1-\nu}{2 \nu}\right)} \, \Gamma\left( \nu \right)^{-\tfrac{1}{\nu}}.
\end{split}
\end{equation}
The expectation values of the descendants of the field $\Phi_{\alpha+2m\tfrac{1-\nu}{\nu}}$ can also be computed from (\ref{expform}) by 
the application of the formula \cite{SGV}:
\begin{equation} \label{mexpform}
\begin{split}
&\beta_{I^+}^* \, \bar{\beta}_{\bar{I}^+}^* \, \bar{\gamma}_{\bar{I}^-}^* \, \gamma_{I^-}^* \, \Phi_{\alpha+2m\tfrac{1-\nu}{\nu}}(0) \cong \\
&C_m(\alpha) \, \beta_{I^++2m}^* \, \bar{\beta}_{\bar{I}^+-2m}^* \, \bar{\gamma}_{\bar{I}^-+2m}^* \, \gamma_{I^--2m}^* \, 
\, \beta_{I_{odd}(m)}^*\, \bar{\gamma}_{I_{odd}(m)}^*\, \Phi_{\alpha}^{(m)}(0), 
\end{split}
\end{equation}
where $\cong$ means identification of operators under expectation value and  for any "vector" $I$ with elements $\{i_j\}$ the vector $I\pm 2m$ denotes a vector with elements $\{i_j\pm 2m\}.$ 
The main formula (\ref{expform}) implies, that actually only ratios of expectation values can be computed in terms of the function $\omega.$ Nevertheless, as a consequence of 
(\ref{primrep}), there is a set of operators whose expectation values can be computed by (\ref{expform}). These operators are nothing but the primaries $\Phi_{2 m\tfrac{1-\nu}{\nu}}$ 
and their descendants, which form the operator content of the sine-Gordon model, when it is defined as a perturbed $c=1$ CFT (See section \ref{sect2b}). 
Here we just note, that under expectation value, we mean a diagonal matrix element of the operator under consideration, such that the sandwiching eigenstate of the Hamiltonian 
is normalized to one.

\section{Large volume checks} \label{sect5}

In this paper we performed 3 important checks in the large volume limit. 
Each check is done in the pure multi-soliton sector of the model. 
The reason for this specialization is that the LeClair-Mussardo type series conjecture of \cite{En} is valid 
only for the solitonic expectation values of the theory. 
The 3 checks are as follows. 
First, we  check an identity for the function $\omega(\alpha)$, which ensures the compatibility of the 
formulas (\ref{expform}) and (\ref{mexpform}) in the large volume limit.
Then, for the operator $\Phi_{4 (1-\nu)/\nu}$ we check upto 3 particle contributions, 
that the solitonic connected form-factors extracted from the ground state expectation value, 
are the ones which enter the  formulas for the 
multi-soliton diagonal form-factors as coefficient functions of the different multi-particle densities. 
Finally, we compute the classical limit of our formula (\ref{expform}) for the multi-soliton expectation values of 
the operators $\Phi_{2 \, n(1-\nu)/\nu}, \quad n\in \mathbb{N},$ and 
from the results, we extract the classical limit of connected diagonal form-factors upto two soliton states. 
The form of these classical connected diagonal form-factors agree with the those coming from direct semi-classical computations in the sine-Gordon 
model \cite{BJsg}.


\subsection{Compatibility check of the formulas (\ref{expform}) and (\ref{mexpform})}

In the original paper \cite{SGV}, it has been shown, that the compatibility of 
formulas (\ref{expform}) and (\ref{mexpform}) require $\omega(\alpha)$ to satisfy 
the following 
identities 
under the analytical continuation of the parameter $\alpha\to \alpha_\pm=\alpha \pm 2 \tfrac{1-\nu}{\nu}:$ 
\begin{equation} \label{omcons}
\begin{split}
\omega_{PQ}(\alpha_\pm)=\frac{\text{sign}(P\pm2) \, \text{sign}(Q\mp2)}{\text{sign}(P) \, \text{sign}(Q)}
\bigg[\omega_{P\pm 2,Q\mp 2}(\alpha)-
\frac{\omega_{P\pm 2,\mp 1}(\alpha)\, \omega_{\pm 1,Q\mp 2}(\alpha)}{\omega_{ \pm 1,\mp 1}(\alpha)+ \tfrac{i}{\nu}
\, \cot\left(\tfrac{\pi}{2 \nu}\pm\tfrac{\pi}{2} \alpha \right)} \bigg].
\end{split}
\end{equation}
This formula is valid for any $P,Q \in {\mathbb Z},$ such that the range of validity in $\alpha$ is 
restricted to the ranges $0<\alpha<2(1-p),$ and $2p<\alpha<2$  for the analytical continuations 
$\alpha \to \alpha_+$ and $\alpha \to \alpha_-,$ respectively. Here, the parameter $p$ is defined by:
\begin{equation} \label{pdef}
\begin{split}
p=\frac{1-\nu}{\nu}, \qquad 0<\nu<1.
\end{split}
\end{equation}

When expectation values in excited states are considered, additional source like terms arise in the 
equations with respect to the ground state description. The equations with these additional source terms 
should respect the identities (\ref{omcons}). Since these source terms are present also in the 
large volume limit, to show that the excited state equations (\ref{omkif}) with (\ref{star2})
 for $\omega(\alpha)$ satisfy (\ref{omcons}) in the infrared limit, is a nontrivial 
test on the excited state formulation of the expectation values. In this part, we show that for multi-soliton 
expectation values, the identities (\ref{omkif}), are satisfied in the infrared limit. 

The first step is to determine $\omega(\alpha)$ in the large volume limit. To do so, equation (\ref{Gj}) with convolution (\ref{star2})
should be solved at large $\ell.$ As a consequence of (\ref{ddv1}), in the $\star$ convolution the integral terms become exponentially 
small in the volume, thus at leading order we neglect them. Thus for large volume and for pure multi-soliton states equation (\ref{Gj}) takes the form: 
\begin{equation} \label{Gjinf}
\begin{split}
{\cal G}_s(\lambda)-2 \pi \, i \, \sum\limits_{j=1}^m \frac{R(\lambda-h_j|\alpha)\, {\cal G}_s(h_j)}{a'(h_j)}=e_{s}(\lambda).
\end{split}
\end{equation}
It contains the discrete values: ${\cal G}_s(h_j), $ which can be determined from the linear equations obtained by taking (\ref{Gjinf}) 
at the positions of the holes:
\begin{equation} \label{Gjhole}
\begin{split}
{\cal G}_s(h_k)-2 \pi \, i \, \sum\limits_{j=1}^m \frac{R(h_k-h_j|\alpha)\, {\cal G}_s(h_j)}{a'(h_j)}=e_{s}(h_k), \qquad k=1,..,m.
\end{split}
\end{equation}
The solution can be written as:
\begin{equation} \label{Gjholesol}
\begin{split}
{\cal G}_s(h_j)=\sum\limits_{k=1}^m a'(h_j) \, \left(\psi^{(\alpha)}(\vec{h})^{-1}\right)_{jk} \, e_s(h_k), \qquad j=1,..,m,
\end{split}
\end{equation}
where the $\alpha$ dependent $m \times m$ matrix $\psi^{(\alpha)}(\vec{h})$ is of the form: 
\begin{equation} \label{psialf}
\begin{split}
\psi^{(\alpha)}_{jk}(\vec{h})=a'(h_j)-2 \, \pi \, i\, R(h_j-h_k|\alpha) \qquad j,k=1,..,m.
\end{split}
\end{equation}
Inserting this into (\ref{Gjinf}) one obtains the large volume solution for ${\cal G}_j(\lambda):$
\begin{equation} \label{Gjinfmo}
\begin{split}
{\cal G}_s(\lambda)=e_s(\lambda)+2 \pi \, i \, \sum\limits_{j,k=1}^m R(\lambda-h_j|\alpha)\, \left(\psi^{(\alpha)}(\vec{h})^{-1}\right)_{jk} \, e_s(h_k). 
\end{split}
\end{equation}
Inserting (\ref{Gjinfmo}) into (\ref{omkif}), one ends up with the following result for the  large volume limit of the matrix $\omega(\alpha):$
\begin{equation} \label{ominfa}
\begin{split}
\omega_{sq}(\alpha)=-\tfrac{2}{\nu^2} \,\text{sign}(s) \, \text{sign}(q) \,  \sum\limits_{j,k=1}^m e_s(h_j) \, \left(\psi^{(\alpha)}(\vec{h})^{-1}\right)_{jk} \, e_q(h_k)+O(e^{-\ell}).
\end{split}
\end{equation}
In this limit, the $\alpha$ dependence of $\omega(\alpha)$ is given by the $\alpha$-dependence of the function $R$ in (\ref{psialf}). We just note, that 
$a'(\lambda)$ is $\alpha$ independent. The key point in the derivation of (\ref{omcons}), is to know, how $\omega(\alpha)$ changes, when the analytical continuation 
$\alpha\to \alpha_\pm=\alpha \pm 2 \tfrac{1-\nu}{\nu}$ is achieved. Since the $\alpha$ dependence is coming from the function $R$ of (\ref{Rfv}), one should know, how it transforms 
under the $\alpha \to \alpha_\pm$ analytical continuation.

From (\ref{Rfv}), it is easy to see, that in the Fourier-space the following relations hold:
\begin{equation} \label{FRfv}
\begin{split}
\tilde{R}(\omega|\alpha_\pm)=\tilde{R}(\omega\pm\tfrac{2 i}{\nu}|\alpha).
\end{split}
\end{equation}
In the $\lambda$-space, it implies the transformations:
\begin{equation} \label{Rapm}
\begin{split}
R(\lambda|\alpha_+)&=e^{ \tfrac{2 \lambda}{\nu}} R(\lambda|\alpha)+\frac{e^{ \tfrac{\lambda}{\nu}}}{\pi \nu} \, \tan\big[\tfrac{\pi}{2 \nu}+ \tfrac{\pi \alpha}{\nu}\big]
+ \sum\limits_{k=1}^{[\alpha/2+p]} \frac{e^{\left(\tfrac{2}{\nu}-\tfrac{2k-\alpha}{1-\nu}\right)\lambda}}{\pi (1-\nu)} \, \tan[\tfrac{\pi \, \nu}{2} \tfrac{2k-\alpha}{1-\nu}], \\
R(\lambda|\alpha_-)&=e^{- \tfrac{2 \lambda}{\nu}} R(\lambda|\alpha)+\frac{e^{- \tfrac{\lambda}{\nu}}}{\pi \nu} \, \tan\big[\tfrac{\pi}{2 \nu}- \tfrac{\pi \alpha}{\nu}\big]
- \!\!\!\!\sum\limits_{k=[\alpha/2-p]+1}^{0} \frac{e^{-\left(\tfrac{2}{\nu}+\tfrac{2k-\alpha}{1-\nu}\right)\lambda}}{\pi (1-\nu)} \, \tan[\tfrac{\pi \, \nu}{2} \tfrac{2k-\alpha}{1-\nu}],
\end{split}
\end{equation}
where $[..]$ stands for integer part in the summation limits. The ranges $0<\alpha<2(1-p),$ and $2p<\alpha<2$ for the continuations 
$\alpha \to \alpha_+$ and $\alpha \to \alpha_-,$ respectively, are chosen to avoid dealing with the last sum in (\ref{Rapm}).
Then, the matrix $\psi^{(\alpha)}(\vec{h})$ at the analytically continued points $\alpha_\pm$ takes the form:
\begin{equation} \label{psiapm}
\begin{split}
\psi^{(\alpha_\pm)}_{jk}(\vec{h})=e_{\pm 2}(h_j) \, \tilde{\psi}^{(\alpha_\pm)}_{jk}(\vec{h}) \, e_{\mp 2}(h_k), \qquad j,k=1,..,m,
\end{split}
\end{equation}
where $e_j(\lambda)$ is given in (\ref{ejfv}) and 
\begin{equation} \label{tpsiapm}
\begin{split}
\tilde{\psi}^{(\alpha_\pm)}_{jk}(\vec{h})={\psi}^{(\alpha)}_{jk}(\vec{h})-C_\nu^\pm \, e_{\mp 1}(h_j) \, e_{\pm 1}(h_k), 
\quad \text{with} \quad C_\nu^\pm=\tfrac{2 i}{\nu}  \tan \left(\tfrac{\pi}{2 \nu}\pm \tfrac{\pi \alpha}{2} \right).
\end{split}
\end{equation}
To express $\omega(\alpha_\pm)$ in terms of $\omega(\alpha)$ from (\ref{ominfa}), 
the inverse matrix of $\tilde{\psi}^{(\alpha_\pm)}_{jk}(\vec{h})$ or equivalently of ${\psi}^{(\alpha_\pm)}_{jk}(\vec{h})$ should be 
expressed in terms of the inverse  of $\tilde{\psi}^{(\alpha)}_{jk}(\vec{h}).$  
The determination of the inverse can be done with the help of the following lemma: 
\newline{\bf Lemma:} \newline
{\it Let $H$ an $m \times m$ invertible matrix, and $e^\pm$ are $m$-dimensional vectors. Then, the inverse of the matrix:
\begin{equation} \label{MH}
\begin{split}
M_{ij}=H_{ij}-C\, e^-_i\, e^+_j, \qquad i,j=1,...,m,
\end{split}
\end{equation}
takes the form as follows:
\begin{equation} \label{MHinv}
\begin{split}
(M^{-1})_{ij}=(H^{-1})_{ij}-\frac{C}{C \, e_r^+  \,(H^{-1})_{rq} \, e^-_q-1} \, 
\left[ (H^{-1})_{is} \, e^-_{s} \right] \,
\left[ e^+_t  (H^{-1})_{tj} \right], \, \quad i,j=1,...,m,
\end{split}
\end{equation}
where for the repeated indexes summation is meant.
}
\newline
The application of the lemma (\ref{MHinv}) to the matrices $\psi^{(\alpha_\pm)}(\vec{h})$ leads to the following formulas for the inverses:
\begin{equation} \label{psiapminv}
\begin{split}
(\psi^{(\alpha_\pm)})^{-1}_{jk}(\vec{h})=e_{\pm 2}(h_j) \, (\tilde{\psi}^{(\alpha_\pm)})^{-1}_{jk}(\vec{h}) \, e_{\mp 2}(h_k), \qquad j,k=1,..,n,
\end{split}
\end{equation}
where 
\begin{equation} \label{tpsiapminv}
\begin{split}
(\tilde{\psi}^{(\alpha_\pm)})^{-1}_{jk}(\vec{h})=({\psi}^{(\alpha)})^{-1}_{jk}(\vec{h})
-\frac{C_\nu^\pm \sum\limits_{l,r=1}^m \left[ ({\psi}^{(\alpha)})^{-1}_{jl}(\vec{h}) \, e_{\mp 1}(h_l) \right]\, \left[ e_{\pm 1}(h_r) \, ({\psi}^{(\alpha)})^{-1}_{rk}(\vec{h}) \right] }
{C_\nu^\pm \sum\limits_{s,q=1}^m e_{\pm 1}(h_s) \, ({\psi}^{(\alpha)})^{-1}_{sq}(\vec{h}) \, e_{\mp 1}(h_q)-1  }.
\end{split}
\end{equation}
Now, $\omega_{PQ}(\alpha_\pm)$ can be computed by inserting (\ref{psiapminv}) into (\ref{ominfa}). 
The special form of the inverse matrix (\ref{psiapminv}) and the identity $e_j(h) \, e_k(h)=e_{j+k}(h), $  
immediately give the identities (\ref{omcons}).



\subsection{Connected diagonal form-factors of $\Phi_{4 \tfrac{1-\nu}{\nu}}(0)$}

In this subsection, we perform a consistency check on the structure of the large volume expansion 
of the expectation value formula (\ref{expform}). This check is based on the conjecture of \cite{En} 
for the large volume series representation of expectation values of local operators in pure multi-soliton 
states. The conjecture was based on the analogous conjecture \cite{LM99,saleur,Pozsg11,Pozsg13,PST14} for purely elastic scattering theories. For the sake of completeness we recall it.
\newline {\bf Conjecture:}
{\it
Let ${\cal O}(x)$ a local operator in the sine-Gordon model. Its expectation value in a pure $n$-soliton state 
with rapidities $\{\theta_1,\theta_2,...,\theta_n\}$ can be written as:
\begin{equation} \label{Ps1}
\begin{split}
\langle \theta_1,...,\theta_n|{\cal O}(x)|\theta_1,...,\theta_n \rangle=&\frac{1}{\hat{\rho}(\theta_1,..,\theta_n)} \\
& \times \sum\limits_{\{\theta_+\}\cup \{\theta_-\}} {\cal D}^{{\cal O}}(\{\theta_+\}) \, \hat{\rho}(\{\theta_-\}|\{\theta_+\}),
\end{split}
\end{equation}
where $\hat{\rho}(\vec{\theta})$ is the determinant of the exact Gaudin-matrix:
\begin{equation} \label{ro}
\hat{\rho}(\theta_1,..,\theta_n)=\text{det} \, \hat{\Phi}(\vec{\theta}), \qquad \hat{\Phi}_{kj}(\vec{\theta})=\frac{d}{d \theta_j} Z(\nu \theta_k|\nu \vec{\theta}), \qquad j,k=1,..,m,
\end{equation}
$Z(\lambda|\vec{h})$ is the counting-function satisfying (\ref{ddv1}) with only hole-type source terms, 
and the sum in (\ref{Ps1}) runs for all bipartite partitions of the rapidities of the sandwiching state:
$\{\theta_1,..,\theta_n \}=\{\theta_+\}\cup\{\theta_-\},$ such that
\begin{equation} \label{ropm}
\hat{\rho}(\{\theta_+\}|\{\theta_-\}=\text{det} \, \hat{\Phi}_+(\vec{\theta}),
\end{equation}
with $\hat{\Phi}_+(\vec{\theta})$ being the sub-matrix of $\hat{\Phi}(\vec{\theta})$ corresponding to the subset $\{\theta_+\}.$
The quantity ${\cal D}^{\cal O}(\{\theta\})$ in (\ref{Ps1}) is called the dressed form-factor \cite{PST14} and
it is given by an infinite sum in terms of the connected diagonal form-factors of the theory:
\begin{equation} \label{ODdress}
\begin{split}
{\cal D}^{\cal O}(\{\theta_1,...,\theta_n\})=\sum\limits_{n_+=0}^\infty \sum\limits_{n_-=0}^\infty 
\frac{1}{n_+! \, n_-!} \! \int\limits_{-\infty}^\infty \! \prod\limits_{i=1}^{n_++n_-} \! \! \frac{d \theta_i}{2 \pi} \!
\prod\limits_{i=1}^{n_+} \! {\cal F}_+(\theta_i+i \, \eta) \! \! \! \prod\limits_{i=n_++1}^{n_++n_-} \! \!
{\cal F}_-(\theta_i-i \, \eta)\\
\times 
F^{\cal O}_c(\theta_1,\theta_2,...,\theta_n,\theta_1\!+\!i \, \eta,...,\theta_{n_+}\!+\!i \, \eta,\theta_{n_++1}\!-\!i \, \eta,...,\theta_{n_++n_-}\!-\!i \, \eta),
\end{split}
\end{equation}
where $F^{{\cal O}}_{c}$ denotes the connected diagonal multi-soliton form factors of ${\cal O}(x),$  \newline
$\eta\in (0,\,\text{min}(p \pi,\pi))$  is a small contour deformation parameter
 and ${\cal F}_\pm(\theta)$ are the nonlinear expressions of the counting function given 
by{\footnote{Here the counting-function is used in the $\lambda$ variable and not in the usual rapidity one. 
The connection is a simple scaling: $\lambda=\nu\,\theta.$}}: 
\begin{equation} \label{calF}
\begin{split}
{\cal F}_{\pm}(\theta)=\frac{(-1)^\delta \, e^{\pm i Z(\nu\, \theta)}}{1+(-1)^\delta \, e^{\pm i Z(\nu\, \theta)}}.
\end{split}
\end{equation}
}
The idea of the consistency check is as follows. As a consequence of the nonlinear integral equations for 
$Z(\lambda)$  (\ref{ddv1}), the functions ${\cal F}_\pm(\theta\pm i \, \eta)$ 
are exponentially small at large volume. Thus the large volume series for a multi-soliton expectation value 
can be rephrased as an expansion in the functions ${\cal F}_\pm(\theta).$ Conjecture (\ref{ODdress}) 
suggests, that the 
coefficient functions in this series are the connected diagonal multi-soliton form factors. 
But, that point of the formula, where a specified connected form-factor enters  
the series, depend on the state in which the expectation value is taken.  
This allows us to make a consistency check on (\ref{expform}). 
In papers \cite{En,En1} the conjecture was tested for the $U(1)$ current and for the trace of the 
stress-energy tensor. Both operators are related to some conserved quantity of the theory. 
To test the conjecture, we choose a simple operator, which is not related to conserved quantities of the 
theory. Our choice is the primary: $\Phi_{4 (1-\nu)/\nu}(0),$ which is the simplest one in the 
series $\Phi_{2 m (1-\nu)/\nu}(0),$ after the trace of the stress-energy tensor 
$ T^{\mu\,}_\mu \sim \Phi_{2 (1-\nu)/\nu}.$
The testing method goes as follows: with the help of (\ref{Ps1}-\ref{ODdress})
from the large volume series of the vacuum expectation value, the $k$-particle 
connected diagonal multi-soliton form-factors can be extracted as 
coefficient functions of the combination ${\cal F}_+(\theta_1)...{\cal F}_+(\theta_k)$.
 We did it upto 3-particle contributions. For the first sight, the method seems ambiguous, since 
everything is valid under integration, but the sought form-factors are symmetric in the rapidities, 
thus after symmetrization in the rapidities, the results for the connected form-factors become unique.  

Having extracted the connected solitonic form-factors from the ground state expectation values, one can check,  
whether the same functions enter the large volume series expansion of 
the expectation values in multi-soliton states. 

From the ground state expectation value, we determine the first three connected diagonal multi-soliton 
form-factors of $\Phi_{4 (1-\nu)/\nu}(0),$ and check, that the large volume expansion of 
the expectation value formula (\ref{expform}) is consistent 
with the form of the Bethe-Yang limit of conjecture (\ref{Ps1}-\ref{ODdress}) 
for 1-, 2- and 3-soliton states. 


To achieve this computation, one should apply (\ref{expform}) for our specific operator 
$\Phi_{4 (1-\nu)/\nu}(0),$ to get a formula for the expectation value. 
To avoid carrying unnecessary constants, we compute connected solitonic form-factors of the 
operator:
\begin{equation} \label{P4bar}
\begin{split}
\bar{\Phi}_{4 \tfrac{1-\nu}{\nu}}(0)=\frac{\Phi_{4 \tfrac{1-\nu}{\nu}}(0)}{C_2(0)\, \mu^{\tfrac{8}{\nu}-8}}.
\end{split}
\end{equation}
According to (\ref{expform}), its expectation value takes the form:
\begin{equation} \label{F4exp}
\begin{split}
\langle \bar{\Phi}_{4 \tfrac{1-\nu}{\nu}}(0) \rangle=\varphi_0 +\varphi_1 \, \omega_{3,-3}(0)
+\varphi_2 \, \omega_{1,-1}(0)+\omega_{1,-1}(0)\, \omega_{3,-3}(0)-\omega_{1,-3}(0)\, \omega_{3,-1}(0),
\end{split}
\end{equation}
where the constants take values:
\begin{equation} \label{varfis}
\begin{split}
\varphi_0=-\tfrac{1}{\nu^2}\, \cot\tfrac{\pi}{2 \nu} \, \cot \tfrac{3 \, \pi}{2 \nu}, \qquad
\varphi_1=\tfrac{i}{\nu}\, \cot\tfrac{\pi}{2 \nu}, \qquad 
\varphi_2=\tfrac{i}{\nu}\, \cot\tfrac{3 \,\pi}{2 \nu}. 
\end{split}
\end{equation}
This implies, that the large volume expansion of $\langle \bar{\Phi}_{4 \tfrac{1-\nu}{\nu}}(0) \rangle$ 
requires the expansion $\omega_{jk}(0)$ in terms of the functions ${\cal F}_{\pm}(\theta).$
The matrix $\omega_{jk}(\alpha)$ is a functional of the function 
${\cal G}_j(\lambda)$ which satisfies (\ref{Gj}). Thus, to get the large volume series 
for an expectation value, the following steps should be taken. From (\ref{Gj}) the large volume 
series for ${\cal G}_j(\lambda)$ should be computed. Then one should insert it into the 
formula (\ref{omkif}) to get the analogous series for $\omega_{jk}(0).$ 
Finally, substituting the large volume series representation of $\omega(0)$ into (\ref{F4exp}), gives 
the required large volume series expression for the expectation value.

Thus, we start the computations by the ground state expectation value, and 
extract the connected solitonic form-factors from the large volume series representation obtained. 
The whole computation is very straightforward. The only subtle point is, that one should treat 
more carefully the integration contour shifts, than it was treated in (\ref{star2}) by a $\pm i 0$ 
prescription in the integration measure. For the ground state ${\cal G}_j(\lambda)$ satisfies the equation:
\begin{equation} \label{GjF}
\begin{split}
{\cal G}_j(\lambda)=e_j(\lambda)-\sum\limits_{\epsilon=\pm}\int\limits_{-\infty}^\infty d\tau 
\, R(\lambda-\tau-i \epsilon \eta|0) \, {\cal F}_\epsilon((\tau+i \epsilon \eta)/\nu).
\end{split}
\end{equation}
To get this equation in the language of ${\cal F}_\pm,$ we used (\ref{calF}) and (\ref{mertek}). 
At large volume, the equations can be solved by a straightforward iterative method. 
The iteration process can be done in the easiest way, by introducing an abstract multi-index notation as follows. For a pair of variables $(\epsilon,\lambda),$ we introduce an abstract capital letter index $A,$ 
such that summation for $A$ unifies summation for $\epsilon$ and integration for $\lambda:$ 
\begin{equation} \label{sumA}
\begin{split}
(\epsilon,\lambda) \to A, \qquad \sum\limits_{\epsilon=\pm}^\infty \int\limits_{-\infty}^\infty \to \sum\limits_{A}. 
\end{split}
\end{equation}
Then it is worth to introduce the "vectors" and "matrices" as follows:
\begin{equation} \label{Avars}
\begin{split}
e_j^{(\epsilon)}(\lambda)=e_{j}(\lambda+i \epsilon \eta)\quad  &\to \quad e_j^{(A)}, \\
R_{\epsilon_1 \epsilon_2}(\lambda_1-\lambda_2)=R(\lambda_1-\lambda_2+i (\epsilon_1-\epsilon_2) \eta|0) \quad &\to \quad R_{A_1 A_2}, \\
\hat{\cal F}_\epsilon(\lambda)={\cal F}_\epsilon((\lambda+i \epsilon \eta)/\nu) \quad &\to \quad {\cal F}_A, \\
{\cal G}_j^{(\epsilon)}(\lambda)={\cal G}_{j}(\lambda+i \epsilon \eta)\quad  &\to \quad {\cal G}^{(A)}_j, \\
Q_{AB}&=R_{AB} \, {\cal F}_B, \qquad \text{(No summation for B)},\\
\check{e}_k^{(A)}&=e_k^{(A)}\, {\cal F}_A. \qquad \text{(No summation for A)}.
\end{split}
\end{equation}
In this language the large volume series for ${\cal G}_j$ and $\omega(0)$ take the form:
\begin{equation} \label{Gjom}
\begin{split}
{\cal G}_j^{(A)}&=(1+Q)^{-1}_{AB}\, e^{(B)}_j, \\
\omega_{jk}(0)&=-\frac{i}{\pi \nu^2} \text{sign}(j)\,\text{sign}(k)\, \check{e}_{j}^{(A)}\, (1+Q)^{-1}_{AC}\, e^{(C)}_k,
\end{split}
\end{equation}
where summation is meant for repeated indexes.
The matrix $Q=O(\cal F),$ thus the power series of $(1+Q)^{-1}$ admits, the required large volume series:
\begin{equation} \label{omQser}
\begin{split}
\omega_{jk}(0)&=-\frac{i}{\pi \nu^2} \text{sign}(j)\,\text{sign}(k)\, \check{e}_{j}^{(A)}\, 
\sum\limits_{n=0}^\infty (-1)^n\,(Q^n)_{AC}\, e^{(C)}_k.
\end{split}
\end{equation}
Inserting (\ref{omQser}) into (\ref{F4exp}) and returning to the original integration variables, 
upto $O({\cal F}^3)$ one ends up with the following result for the ground state expectation value: 
\begin{equation} \label{expgs0}
\begin{split}
&\langle \bar{\Phi}_{4 \tfrac{1-\nu}{\nu}}(0) \rangle_0\!=\!\!\!\sum\limits_{\epsilon_1=\pm}\! \int\limits_{-\infty}^{\infty} 
\frac{d\tau}{2 \pi} \, \hat{\cal F}_{\epsilon_1}(\tau)\, \bar{F}_c^{(1)}(\tau^{(\epsilon_1)})\!+\!\!\!\!\!\!\!
\sum\limits_{\epsilon_1,\epsilon_2=\pm} \int\limits_{-\infty}^{\infty} \!\!\!
\frac{d\tau \,d\tau'}{(2\pi)^2 } \hat{\cal F}_{\epsilon_1}(\tau) \hat{\cal F}_{\epsilon_2}(\tau') 
 \frac{\bar{F}_c^{(2)}(\tau^{(\epsilon_1)}\!,\tau'^{(\epsilon_2)})}{n_{\epsilon_1,\epsilon_2}}\\
&+\sum\limits_{\epsilon_1,\epsilon_2,\epsilon_3=\pm} \int\limits_{-\infty}^{\infty} 
\frac{d\tau \,d\tau'\, d\tau''}{(2 \pi)^3} \hat{\cal F}_{\epsilon_1}(\tau) \hat{\cal F}_{\epsilon_2}(\tau')\,\hat{\cal F}_{\epsilon_3}(\tau'')\,
 \frac{\bar{F}_c^{(3)}(\tau^{(\epsilon_1)},\tau'^{(\epsilon_2)},\tau''^{(\epsilon_3)})}{n_{\epsilon_1,\epsilon_2,\epsilon_3}}+O({\cal F}^4),
\end{split}
\end{equation}
where we introduced the short notation $\tau^{(\epsilon)}=\tau+i \epsilon \eta.$ The symmetry factors 
are given as a product of two factorials: $n_{\epsilon_1,..}=(\#\text{ of + indexes})!\,(\# \text{ of - indexes})!.$ In the "$\lambda$-convention", 
the connected diagonal form-factors of $\bar{\Phi_{4(1-\nu)/\nu}}(0)$ take the form:
\begin{equation} \label{barFcs}
\begin{split}
&\bar{F}_c^{(1)}(\tau)=\tfrac{2 i}{ \nu^2}\, (\varphi_1+\varphi_2), \\
&\bar{F}_c^{(2)}(\tau,\tau')=-\tfrac{4 \pi \,i}{ \nu^2}\,  \varphi_1 \,
 \left(e_3(\tau)\, e_{-3}(\tau') +e_{-3}(\tau)\, e_{3}(\tau') \right) \, R(\tau-\tau'|0) \\
&-\tfrac{4  \pi \, i}{ \nu^2}  \varphi_2 \! 
\left(e_1(\tau) e_{-1}(\tau') \!+\!e_{-1}(\tau) e_{1}(\tau') \right) \! R(\tau-\tau'|0)\!+\!
\tfrac{4 }{\nu^4}\left(e_2(\tau) e_{-2}(\tau')\!+\!e_{-2}(\tau) e_{2}(\tau')\!-\!2 \right),
\end{split}
\end{equation}
\begin{equation} \label{barFcs3}
\begin{split}
&\bar{F}_c^{(3)}(\tau_1,\tau_2,\tau_3)=(2 \pi)^3\sum\limits_{\sigma \in S_3} \bigg\{ 
\tfrac{i}{\pi \, \nu^2}\,  \varphi_1 \,  e_3(\tau_{\sigma(1)}) R_{\sigma(1),\sigma(2)} \,
R_{\sigma(2),\sigma(3)} \, e_{-3}(\tau_{\sigma(3)})\\
&+\tfrac{i}{\pi \, \nu^2}\,  \varphi_2 \,  e_1(\tau_{\sigma(1)}) R_{\sigma(1),\sigma(2)} \,
R_{\sigma(2),\sigma(3)} \, e_{-1}(\tau_{\sigma(3)})
+\tfrac{1}{\pi^2 \nu^4} \bigg[
 e_1(\tau_{\sigma(1)}) R_{\sigma(1),\sigma(2)} \, e_{-1}(\tau_{\sigma(2)})\\
&+e_3(\tau_{\sigma(2)}) R_{\sigma(2),\sigma(3)} \, e_{-3}(\tau_{\sigma(3)})-
e_1(\tau_{\sigma(1)}) R_{\sigma(1),\sigma(2)} \, e_{-3}(\tau_{\sigma(2)}) \, e_{2}(\tau_{\sigma(3)})\\
&-e_{-2}(\tau_{\sigma(1)}) \,e_{3}(\tau_{\sigma(2)})R_{\sigma(2),\sigma(3)} \, e_{-1}(\tau_{\sigma(3)}) \, 
\bigg] \bigg\},
\end{split}
\end{equation}
where in the expression of $\bar{F}_c^{(3)},$ the sum runs for all permutations of the 3 indexes,
and we introduced the short notation $R_{jk}=R(\tau_j-\tau_k|0).$

Now, one can compute the large volume limit of the expectation values from (\ref{expform}) 
for multi-soliton states. 
From (\ref{Gjinfmo}) and (\ref{ominfa}), for $\omega(0),$ the result as follows is obtained: 
\begin{equation} \label{ominf}
\begin{split}
\omega_{jk}^{BY}(0)=-\frac{2 \,i}{\nu^2} \, \text{sign}(j) \, \text{sign}(k) \, \sum\limits_{s,q=1}^{m} 
e_{j}(h_s) \, \Phi^{-1}_{sq}(h) \, e_{k}(h_q),
\end{split}
\end{equation}
where $e_j(h)$ is defined in (\ref{ejfv}) and the Gaudin-matrix is given by:
\begin{equation} \label{Gauinf}
\begin{split}
&\Phi_{jk}(h)=\left\{ 
\begin{array}{r}
\rho_1(h_j)-2 \pi \sum\limits_{s\neq j}^m R(h_j-h_s|0), \qquad \qquad \qquad \qquad
\qquad j=k, \\
2 \pi \, R(h_j-h_k|0), \qquad \qquad \qquad \qquad  \qquad j\neq k,
\end{array}\right. \\
&\text{with} \qquad \qquad \qquad
\rho_1(h)=\tfrac{\ell}{\nu} \, \cosh \tfrac{h}{\nu}.
\end{split}
\end{equation}
The positions of the holes in this formula are related to the rapidity variables of (\ref{Ps1}) by $\tfrac{h_j}{\nu}=\theta_j.$ 
We note, that here $\Phi_{jk}$ is the Gaudin-matrix in the $\lambda$ variable, which is related to that of in rapidity 
variable by a scaling with $\nu:$ 
$\Phi_{jk}(\nu\, \theta_1,...)=\tfrac{1}{\nu} \hat{\Phi}_{jk}(\theta_1,...).$ 
As a consequence, the corresponding $n$-particle densities are related by 
a scaling with 
$\nu^n: \quad \rho^{(n)}(\nu \theta_1,\nu \theta_1,...)=\tfrac{1}{\nu^n}\hat\rho^{(n)}(\theta_1,\theta_2,...).$

From inserting (\ref{ominf}) into (\ref{F4exp}), the large volume limit of the expectation values  of $\bar{\Phi}_{4 \tfrac{1-\nu}{\nu}}(0),$ in 1-, 2-, and 3-soliton states 
can be determined. For 1-particle the result takes the form: 
\begin{equation} \label{1partFc}
\begin{split}
\langle h_1|\bar{\Phi}_{4 \tfrac{1-\nu}{\nu}}(0)|h_1 \rangle=\varphi_0+\frac{1}{\text{det} \Phi(h_1)}\, \bar{F}_c^{(1)}(h_1).
\end{split}
\end{equation}
For the  2-particle expectation value, one obtains:
\begin{equation} \label{2partFc}
\begin{split}
\langle h_1,h_2|\bar{\Phi}_{4 \tfrac{1-\nu}{\nu}}(0)|h_1,h_2 \rangle=\varphi_0+
\frac{\bar{F}_c^{(1)}(h_1) \, \rho^{(1)}(h_2)+\bar{F}_c^{(1)}(h_2) \, \rho^{(1)}(h_1)+\bar{F}_c^{(2)}(h_1,h_2)}{\text{det} \Phi(h_1,h_2)},
\end{split}
\end{equation}
where $\rho^{(1)}(h_j)=\tfrac{\ell}{\nu} \cosh \tfrac{h_j}{\nu}-2 \pi \, R(h_1-h_2|0), \qquad j=1,2.$ \newline
For the 3-particle case the result is as follows:
\begin{equation} \label{3partFc}
\begin{split}
\langle h_1,h_2,h_3|\bar{\Phi}_{4 \tfrac{1-\nu}{\nu}}(0)|h_1,h_2,h_3 \rangle=\varphi_0+\frac{1}{{\text{det} \Phi(h_1,h_2,h_3)}} \, \bigg\{
\bar{F}_c^{(1)}(h_1) \, \rho^{(2)}_3(h_2,h_3|h_1) \\
+\bar{F}_c^{(1)}(h_2) \, \rho^{(2)}_3(h_1,h_3|h_2)+
\bar{F}_c^{(1)}(h_3) \, \rho^{(2)}_3(h_1,h_2|h_3)+
\bar{F}_c^{(2)}(h_1,h_2) \, \rho^{(1)}_3(h_3|h_1,h_2)\\
+\bar{F}_c^{(2)}(h_1,h_3) \, \rho^{(1)}_3(h_2|h_1,h_3)+
\bar{F}_c^{(2)}(h_2,h_3) \, \rho^{(1)}_3(h_1|h_2,h_3)+
\bar{F}_c^{(3)}(h_1,h_2,h_3)
\bigg\},
\end{split}
\end{equation}
where the densities take the forms:
\begin{equation} \label{densities}
\begin{split}
\rho^{(1)}_3(h_j|h_k,h_l)&=\rho_j-\hat{R}_{jk}-\hat{R}_{jl}, \\
\rho^{(2)}_3(h_k,h_l|h_j)&=\rho_k \rho_l-\rho_k (\hat{R}_{kl}+\hat{R}_{lj})-
\rho_l (\hat{R}_{kl}+\hat{R}_{kj})+
\hat{R}_{kl} \hat{R}_{lj}+\hat{R}_{kl} \hat{R}_{kj}+\hat{R}_{kj}\hat{R}_{lj},
\\
\text{with}& \qquad \rho_j=\tfrac{\ell}{\nu} \cosh \tfrac{h_j}{\nu}, \quad 
\text{and} \quad \hat{R}_{jk}=2 \pi \, R(h_j-h_k|0).
\end{split}
\end{equation}
The results (\ref{1partFc})-(\ref{3partFc}) are in perfect agreement with the conjecture (\ref{Ps1}). 

To summarize the results, we close this section by listing the multi-soliton connected diagonal form-factors of the operator $\Phi_{4 \tfrac{1-\nu}{\nu}}(0)$ in rapidity variables. 
They take the form as follows: 
\begin{equation} \label{multsolFF}
\begin{split}
F_c^{(0)}(\emptyset)&=C_2(0)\, \mu^{\tfrac{8}{\nu}-8} \,\varphi_0, \\
F_c^{(1)}(\theta_1)&=C_2(0)\, \mu^{\tfrac{8}{\nu}-8} \, \nu \, \bar{F}_c^{(1)}(\nu \theta_1), \\
F_c^{(2)}(\theta_1,\theta_2)&=C_2(0)\, \mu^{\tfrac{8}{\nu}-8} \, \nu^2 \,\bar{F}_c^{(2)}(\nu \theta_1,\nu \theta_2), \\
F_c^{(3)}(\theta_1,\theta_2,\theta_3)&=C_2(0)\, \mu^{\tfrac{8}{\nu}-8} \, \nu^3 \,\bar{F}_c^{(3)}(\nu \theta_1,\nu \theta_2,\nu \theta_3),
\end{split}
\end{equation}
with the functions $\bar{F}_c(h_1,...)$ given in (\ref{barFcs}) and (\ref{barFcs3}). 
Here the prefactor in front of the functions $\bar{F}_c$ comes from either the proper normalization factor of the operator $\Phi_{4 \tfrac{1-\nu}{\nu}}(0)$
and the transformation from the $\lambda$-type variables to the rapidity 
variables.


\subsection{Classical limit of connected diagonal form factors}

In this subsection, starting from the formulas (\ref{primrep}) and (\ref{expform}), we compute the classical 
limit of the multi-soliton connected diagonal form-factors of the operators $\Phi_{2 n\tfrac{1-\nu}{\nu}}(0)$ 
and compare the results with the direct semi-classical computation of reference \cite{BJsg}. 

The connected diagonal form-factors can be extracted from the expectation values with the help of the 
Bethe-Yang limit of the conjectured expectation value formula (\ref{Ps1})-(\ref{calF}).  

For pure soliton states, in the Bethe-Yang limit the expectation values take exactly the same form 
as those in a purely elastic scattering theory \cite{saleur,PT08b,BWu}. This is in accordance with the earlier conjecture of \cite{Palmai13}, for 
the Bethe-Yang limit of expectation values in non-diagonally scattering theories.

For an $m$-soliton  state with rapidities $\{\theta_1,\theta_2,...,\theta_m\}$ the large volume limit of an expectation value takes the form: 
\begin{equation} \label{Ps11}
\begin{split}
\langle \theta_1,...,\theta_m|{\cal O}(x)|\theta_1,...,\theta_m \rangle=&\frac{1}{\rho(\theta_1,..,\theta_m)} \\
& \times \sum\limits_{\{\theta_+\}\cup \{\theta_-\}} {F}_c^{{\cal O}}(\{\theta_+\}) \, \rho(\{\theta_-\}|\{\theta_+\}),
\end{split}
\end{equation}
where again the sum in (\ref{Ps11}) runs for all bipartite partitions of the rapidities of the sandwiching 
state: $\{\theta_1,..,\theta_n \}=\{\theta_+\}\cup\{\theta_-\},$ such that the partial densities are 
made out of the infinite volume Gaudin-matrix of (\ref{Gauinf}) by the formula;
\begin{equation} \label{ropm}
\rho(\{\theta_+\}|\{\theta_-\}=\text{det} \, {\Phi}_+(\vec{\theta}),
\end{equation}
with ${\Phi}_+(\vec{\theta})$ being the sub-matrix of ${\Phi}(\vec{\theta})$ corresponding to the subset $\{\theta_+\}.$ Now, the coefficients of the densities are directly the connected diagonal form-factors ${F}_c^{\cal O}(\{\theta\})$ .


The strategy of computing the classical limit of the connected diagonal form-factors upto two 
soliton-states is as follows. From (\ref{primrep}) and (\ref{expform}), one can write down explicit formulas for 
the expectation values of ${\cal O}_n=\Phi_{2 n \tfrac{1-\nu}{\nu}}$ in multi-soliton states. 
First, one should take the vacuum expectation value. Then from (\ref{Ps11}) at $m=0,$ one obtains
 $F_{0,c}^{{\cal O}_n}(\{\emptyset\}).$ 
Then by comparing the 1-particle expectation value and (\ref{Ps11}) applied for $m=1,$ and by knowing 
$F_{0,c}^{{\cal O}_n}(\{\emptyset\}),$ one can extract the 1-particle connected form-factors: 
$F_{1,c}^{{\cal O}_n}(\{\theta\}).$  
As a last step, the 2-particle expectation value should be compared to (\ref{Ps11}) at $m=2,$ and by knowing 
$F_{0,c}^{{\cal O}_n}(\{\emptyset\}),$ and $F_{1,c}^{{\cal O}_n}(\{\theta\}),$ one can extract the 2-particle connected form-factors: $F_{2,c}^{{\cal O}_n}(\{\theta_1,\theta_2\}).$ Finally, the classical limit should be 
taken in the resulting connected form-factors.

 For general values of $n,$ and for more than 2 soliton states, 
the computation described above would be very complicated to achieve. 
Nevertheless, as we will see in the case of two-soliton expectation values, the whole computation can be 
simplified, if the classical limit is taken appropriately in the internal steps of the computation.

 To treat the classical limit appropriately, as a first step, we collect some important formulas 
concerning the classical limit of some basic building blocks of the computations. 
The classical limit is the $\nu \to 1$ or equivalently the $p \to 0$ limit of the theory. 
In this limit both the normalization factor $C_n(\alpha)$ in (\ref{Cm}) and the kernel $R(\lambda|0)$ 
given in (\ref{Rfv}) diverge:
\begin{equation} \label{CRclass}
\begin{split}
C_{n}(0)&=\frac{1}{(\pi\, p)^n}\, C_n^{class}(0)+...,\qquad \text{with} \\
C_{n}^{class}(0)&=\frac{(2\,i)^n}{\prod\limits_{j=1}^n (2 j-1)}, \\
2 \pi \, R(\lambda|0)&=\frac{-1}{\pi \, p} \left( 2 \ln \tanh \tfrac{\lambda}{2}+...\right).
\end{split}
\end{equation}
The expectation value of the operators ${\cal O}_n(0)$ requires the matrix $\omega_{jk}(0)$ 
of (\ref{ommatr}) in the $\ell \to \infty$ limit. This is given in formula (\ref{ominf}). 

The application of the formulas (\ref{primrep}) and (\ref{expform}) gives the following representation for the 
Bethe-Yang limit of expectation values of the operators ${\cal O}_n=\Phi_{2 n \tfrac{1-\nu}{\nu}}:$ 
\begin{equation} \label{BYrep}
\begin{split}
\langle O_n(0) \rangle=C_n(0) \, \text{det}_{n \times n}  \, M^{(quant)},
\end{split}
\end{equation}
where $M^{(quant)}$ denotes the Bethe-Yang limit of $\Omega_{jk}$ in (\ref{DR}): 
\begin{equation} \label{Mqu}
\begin{split}
M^{(quant)}_{jk}=\omega_{2j-1,1-2k}^{BY}+\tfrac{i}{\nu} \, \cot\left[ \tfrac{\pi}{2 \nu} (2j-1)\right] \, \delta_{jk}, 
\qquad j,k=1,...,n,
\end{split}
\end{equation}
Using this formula, the vacuum expectation value can be computed with ease. In this case there are no 
particles present, thus $\omega^{BY}=0,$ allowing to compute the necessary determinant easily:  
\begin{equation} \label{expcl0}
\begin{split}
\langle {\cal O}_n(0)\rangle_0=C_n(0)\, \left(\tfrac{i}{\nu} \right)^n \, \prod\limits_{j=1}^n 
\cot\left[ \tfrac{\pi}{2 \nu} (2j-1)\right] \stackrel{p \to 0}{=}1+O(\pi \, p).
\end{split}
\end{equation}
Comparing (\ref{expcl0}) and (\ref{Ps11}) gives the classical limit of the 0-particle connected form-factors: 
\begin{equation} \label{Fczero}
\begin{split}
F_{c,class}^{{\cal O}_n}(\{\emptyset\})=1.
\end{split}
\end{equation}

The computation of the 1-soliton connected form-factor goes as follows. 
Let $A$ a diagonal matrix:
\begin{equation} \label{Adef}
\begin{split}
A_{jk}=A_j \, \delta_{jk}, \qquad A_j=\tfrac{i}{\nu} \, \cot\left[ \tfrac{\pi}{2 \nu}(2j-1)\right], 
\qquad j,k=1,...,n. 
\end{split}
\end{equation}
In this case the matrix  
 $\omega^{BY}_{2j-1,1-2k}$ takes the form of a dyadic  product:
\begin{equation} \label{ominf1}
\begin{split}
\omega^{BY}_{2j-1,1-2k}&=\frac{2i}{\nu^2}\, \frac{1}{\rho_1(h)} \, e^+_{j}(h) \, e^-_{k}(h), \qquad j,k=1,..,n,\\
e_{j}^\pm(h)&=e^{\pm \tfrac{2 j}{\nu}h}, \qquad \qquad \qquad \qquad \qquad j=1,..,n,
\end{split}
\end{equation}
where $\rho_1(h)$ is given in (\ref{Gauinf}). 
The matrix $M^{(quant)}$ takes the special form:
\begin{equation} \label{Mqu1}
\begin{split}
M^{(quant)}=A+\lambda_0 \, \, e^+  {\tiny{\text{O}}} \,e^-, \qquad \text{with} \quad \lambda_0=\frac{2i}{\nu^2} \frac{1}{\rho_1(h)},
\end{split}
\end{equation}
where ${\tiny{\text{O}}}$ stands for the dyadic product. The determinant of such a matrix 
is given by the simple formula:
\begin{equation} \label{detM1}
\begin{split}
\text{det} \, (A+\lambda_0 \, \, e^+  {\tiny{\text{O}}} \,e^-)=
\text{det}\, A \cdot (1+\lambda_0\, \, e^+ A^{-1} e^- ).
\end{split}
\end{equation}
With the help of this formula, one obtains the result as follows for the 1-soliton expectation value,
\begin{equation} \label{exps1}
\begin{split}
\langle {\cal O}_n(0) \rangle_1=F^{{\cal O}_n}_{0,c}(\emptyset)+\frac{F^{{\cal O}_n}_{0,c}(\emptyset)}{\rho_1(h)}\,
\left(\frac{2i}{\nu^2} \sum\limits_{j=1}^n \frac{1}{A_j} \right), 
\end{split}
\end{equation}
where we exploited the previous result (\ref{expcl0}) for the vacuum expectation value: $F^{{\cal O}_n}_{0,c}(\emptyset)=C_n(0)\, \text{det}\, A.$ Comparing (\ref{exps1}) to (\ref{Ps11}) applied for $m=1,$ 
the 1-particle connected diagonal form-factor can be obtained:
\begin{equation} \label{Fc1}
\begin{split}
F_{1,c}^{{\cal O}_n}(\theta)=F^{{\cal O}_n}_{0,c}(\emptyset) \, \left(\frac{2i}{\nu^2} \sum\limits_{j=1}^n \frac{1}{A_j} \right) \stackrel{p \to 0}{=} -\frac{1}{\pi \, p} F^{{\cal O}_n,(0)}_{c,class}(\emptyset) \, 
4 \, \sum\limits_{j=1}^n \frac{1}{2j-1}.
\end{split}
\end{equation}
This result agrees with that of the classical computations \cite{BJsg} for the 1-particle solitonic connected diagonal form-factors. From (\ref{Fc1}), it can bee seen that the 1-particle connected form-factors 
of ${\cal O}_n(0)$ scale as $\frac{1}{\pi \, p}$ in the classical limit. 

As the number of sandwiching solitons increase, it is helpful for the computations to know,
how the connected form-factors scale in the classical 
limit. As a consequence of the scaling (\ref{CRclass}) of the function $R(\lambda|0),$ the multi-soliton 
connected form-factors admit the following scaling property in the classical limit{\footnote{This 
scaling property can be seen in two ways. First, for fixed low values of $n,$ one can do the computations 
such that the classical limit is taken only at the end.  And for each case this scaling is obtained. 
(E.g. for $n=2,$ the quantum result can be found in the previous subsection.)  
Second, the solitonic connected diagonal  form-factors of ${\cal O}_1(0)$ are known explicitly, since this operator is proportional to  the 
trace of the stress-energy tensor. They are composed of simple products of the function $R(\lambda|0)$, 
which allows one to compute easily the classical limit.}}:
\begin{equation} \label{Fcscal}
\begin{split}
F_{m,c}^{{\cal O}_n}(\theta_1,...,\theta_m)\stackrel{p \to 0}{=}\frac{1}{(\pi \,p)^m} \, F_{c,class}^{{\cal O}_n,(m)}(\theta_1,...,\theta_m)+O((\pi \ p)^{1-m}),
\end{split}
\end{equation}
where $F_{c,class}^{{\cal O}_n,(m)}(\theta_1,...,\theta_m)$ is what 
we call now the classical limit of connected form 
factors. Beyond the level of 1-particle expectation values, the computations become so complicated, that 
to simplify them, it is worth to exploit this scaling property at the internal steps of the computations. 

For the 2-soliton expectation values, as a first step we write down the densities appearing in (\ref{Ps11}). 
We consider (\ref{Ps11}) as the function of particle's rapidities and the corresponding 1-particle densities. 
Thus, we write down the Bethe-Yang limit of the 2-particle Gaudin-matrix in this spirit:
\begin{equation} \label{Gau2}
\begin{split}
\hat{\Phi}(\theta_1,\theta_2)=\begin{pmatrix} \rho_1(\nu \theta_1)-2 \pi\, \nu \, R(\nu (\theta_1-\theta_2)|0) &  2 \pi\, \nu \, R(\nu (\theta_1-\theta_2)|0) \\
2 \pi\, \nu \, R(\nu (\theta_1-\theta_2)|0) & \rho_1(\nu \theta_2)-2 \pi\, \nu \, R(\nu (\theta_1-\theta_2)|0) 
\end{pmatrix}.
\end{split}
\end{equation}
From (\ref{CRclass}) and (\ref{Gauinf}), one can see, that $\rho_1(\theta) \sim O(1),$ while 
$R(\nu \theta|0)\sim O(\tfrac{1}{\pi \, p})$ in the classical limit. Thus for computational reasons it is worth to rewrite $\hat{\Phi}(\theta_1,\theta_2)$ in terms of a scaled 1-particle density:
\begin{equation} \label{scrho}
\begin{split}
\rho_c(\theta)=\pi p \, \rho_1(\nu\, \theta).
\end{split}
\end{equation}
One can write down (\ref{Ps11}) in terms of the rapidities and the new densities $\rho_c(\theta)$ and 
take the $p \to 0$ classical limit such that $\rho_c(\theta)$ and the rapidities are kept finite. 
In the classical limit the Gaudin-matrix will behave as follows:
\begin{equation} \label{GauCl}
\begin{split}
\hat{\Phi}(\theta_1,\theta_2)&\stackrel{p \to 0}{=}\frac{1}{\pi \,p} \,\hat{\Phi}^{cl}(\theta_1,\theta_2)+...,\\
\hat{\Phi}^{cl}(\theta_1,\theta_2)&=\begin{pmatrix} \rho_c(\theta_1)+2 \ln u &  -2 \ln u \\
-2 \ln u & \rho_c(\theta_2)+2 \ln u 
\end{pmatrix}, 
 \qquad u=\tanh \tfrac{\theta_1-\theta_2}{2}.
\end{split}
\end{equation}
Then the densities entering (\ref{Ps11}) take the form:
\begin{equation} \label{denses}
\begin{split}
\rho_2(\theta_1,\theta_2)&=\frac{1}{(\pi  p)^2}\, \text{det} \hat{\Phi}^{cl}(\theta_1,\theta_1)+...,\\
\rho(\theta_1|\theta_2)&=\frac{1}{\pi p} \, (\rho_c(\theta_1)+2 \ln u)+..., \\
\rho(\theta_2|\theta_1)&=\frac{1}{\pi p} \, (\rho_c(\theta_2)+2 \ln u)+....
\end{split}
\end{equation}
Inserting (\ref{denses}) and the scaling property (\ref{Fcscal}) into (\ref{Ps11}), the expectation value 
can be written in the $p \to 0$ limit as follows: 
\begin{equation} \label{expcl2}
\begin{split}
\langle {\cal O}_n(0)\rangle_2 \stackrel{p \to 0}{=}&F_{c,class}^{{\cal O}_n,(0)}(\emptyset)
+\frac{1}{\text{det} \hat{\Phi}^{cl}(\theta_1,\theta_2)}
\big( F_{c,class}^{{\cal O}_n,(2)}(\theta_1,\theta_2)+
F_{c,class}^{{\cal O}_n,(1)}(\theta_1) \, \left(\rho_c(\theta_1)+2 \ln u\right)+\\
&F_{c,class}^{{\cal O}_n,(1)}(\theta_2) \, \left(\rho_c(\theta_2)+2 \ln u\right)
\big)+...,
\end{split}
\end{equation}
With the help of (\ref{expcl2}) the extraction of the classical 2-soliton connected form-factor 
goes as follows. The expectation value $\langle {\cal O}_n(0)\rangle_2$ is considered as a function 
of many parameters: 
\begin{equation} \label{interpret1}
\begin{split}
\langle {\cal O}_n(0)\rangle_2={\bf F}(\rho_{c,1},\rho_{c,2},u|p), \quad \text{where} \quad \rho_{c,j}=\rho_{c}(\theta_j).
\end{split}
\end{equation}
(\ref{GauCl}) suggests a similar reparameterization for the classical 2-particle density:
\begin{equation} \label{2denscl}
\begin{split}
\text{det} \hat{\Phi}^{cl}(\theta_1,\theta_2)= \rho_2(\rho_{c,1},\rho_{c,2},u|p).
\end{split}
\end{equation}
Formula (\ref{expcl2}) implies , that the classical 2-soliton connected form-factor can be 
obtained as follows:
\begin{equation} \label{Fc2get}
\begin{split}
F_{c,class}^{{\cal O}_n,(2)}(\theta_1,\theta_2)=\lim\limits_{\rho \to 0}{\bf F}(\rho,\rho,u|0)\, \rho_2(\rho,\rho,u|0)-
4 \, F_{c,class}^{{\cal O}_n,(1)}(\theta_1) \, \ln u.
\end{split}
\end{equation}
We just note, that we exploited, that the 1-particle connected form-factor is independent of the 
rapidity. 
Finally formula (\ref{2denscl}) together with (\ref{GauCl}), (\ref{BYrep}),(\ref{ominf}), (\ref{Mqu}) 
admit the following representation for the 2-soliton diagonal connected form-factors in the variable 
$u=\tanh \tfrac{\theta_1-\theta_2}{2}$: 
\begin{equation} \label{2solFF}
\begin{split}
F_{c,class}^{{\cal O}_n,(2)}(u)=C_n^{class}(0) \,\lim\limits_{\rho \to 0} \text{det}
 \, M^{(class)}(u|\rho) \, \text{det} \, \Phi^{cl}(u|\rho)-
4 \, F_{c,class}^{{\cal O}_n,(1)}(\theta_1) \, \ln u,
\end{split}
\end{equation}
where the matrix $\Phi^{cl}(u|\rho)$ is equal to $\hat{\Phi}^{cl}$ of (\ref{GauCl}) taken at  $\rho_{c,1}=\rho_{c,2}=\rho:$
\begin{equation} \label{gaucl}
\begin{split}
\Phi^{cl}(u|\rho)&=\begin{pmatrix} \rho+2 \ln u &  -2 \ln u \\
-2 \ln u & \rho+2 \ln u 
\end{pmatrix}, 
 \qquad u=\tanh \tfrac{\theta_1-\theta_2}{2},
\end{split}
\end{equation}
the matrix $M^{(class)}(u|\rho)$ can be considered as the classical limit of $M^{(quant)}$ of (\ref{Mqu}). 
It is expressed i term of the classical limit of $\omega^{BY}_{jk}:$
\begin{equation} \label{omBYcl}
\begin{split}
\omega^{BY}_{jk}(u|\rho)=-2 \, i \, \text{sign}(j) \, \text{sign}(k) \, 
\big\{ t(u)^{j+k}\, \Phi^{cl,-1}_{11}+ \Phi^{cl,-1}_{22}+
t(u)^{j}\, \Phi^{cl,-1}_{12}+t(u)^{k}\, \Phi^{cl,-1}_{21}
\big\},
\end{split}
\end{equation}
where $t(u)= \frac{u+1}{u-1}$ and $\Phi^{cl,-1}_{jk}$ denotes the $jk$ matrix element of the 
inverse of $\Phi^{cl}(u|\rho)$ of (\ref{gaucl}). 
Then $M^{(class}(u|\rho)$ is given by the formula: 
\begin{equation} \label{Mcl}
\begin{split}
M^{(class)}(u|\rho)=\omega_{2j-1,1-2k}^{BY}(u|\rho)-\tfrac{i}{2} (2j-1) \, \delta_{jk}, \qquad j,k=1,..,n.
\end{split}
\end{equation}
For the first few values of $n,$ we just list the classical limit of connected 
diagonal 2-soliton form factors:
\begin{equation} \label{fII}
\begin{split}
\frac{F_{c,class}^{{\cal O}_n,(2)}(u)}{F^{{\cal O}_n,(0)}_{c,class}(\emptyset)}\!\!=\!\!\left\{ 
\begin{array}{r}
\frac{16 \left(u^2+1\right) \log
   (u)}{u^2-1}, \qquad  n=1, \\ 
-\frac{256 u^2}{3
   \left(u^2-1\right)^2}+\frac{64 \left(u^2+1\right)^3 \log
   (u)}{3 \left(u^2-1\right)^3}, \qquad  n=2, \\ 
-\frac{512 u^2 \left(3 u^4+2
   u^2+3\right)}{5
   \left(u^2-1\right)^4}+\frac{16 \left(u^2+1\right)
   \left(23 u^8+132 u^6+458
   u^4+132 u^2+23\right) \log
   (u)}{15 \left(u^2-1\right)^5},  \qquad n=3, \\ 
\text{etc....} 
\end{array}\right.
\end{split}
\end{equation}

We compared these results upto $n=8$ with the ones obtained by semi-classical commutations in \cite{BJsg}, and 
found perfect agreement{\footnote{The semi-classical results of Bajnok and Janik can be found in \cite{BJsg} upto $n=4.$ For higher values of $n,$ 
we could use the unpublished Mathematica notebook file of the authors.}}.


\section{Small volume checks} \label{sect6}

In this section in the pure multi-soliton sector, we compare our results to 3-point functions of the Liouville theory on the cylinder.  
Let $\{h_j\}_{j=1}^n$ the rapidities of the pure multi-soliton state. We consider the ultraviolet limit of the expectation value 
$\frac{\langle h_1,..,h_n| {\cal O}(0)| h_1,...h_n\rangle}{\langle h_1,...,h_n|h_1,...,h_n\rangle}.$
The operator under consideration is a primary field or a Virasoro descendant of a primary field:
${\cal O}(0) \to {\bf l}_{-k_1}...{\bf l}_{-k_j} \Phi_\alpha(0),$ and the sandwiching state tends to a primary state or a descendant state of the theory:
$| h_1,...h_n\rangle \to L_{-n_1}...L_{-n_i} |\Delta\rangle.$ 
If we introduce the complex coordinate on the cylinder of radius one as $z=x+i \, y,$ such that $y\equiv y+2\pi, $
then in the previous lines ${\bf l}_n$ and $L_n$ are the coefficients of the Laurent-expansion of 
the stress-energy tensor 
$T(z)$ around the origin and $z \to \pm \infty,$ respectively:
\begin{equation} \label{bfl}
\begin{split}
T(z)=\sum\limits_{n=-\infty}^{\infty} {\bf l}_n z^{-n-2}, \qquad T(z)=\sum\limits_{n=-\infty}^{\infty} L_n e^{nz}-\tfrac{c}{24}.
\end{split}
\end{equation}

The PCFT formulation of the sine-Grodon model, suggests that the following formula should hold 
in the ultraviolet limit: 
\begin{equation} \label{numcomp}
\begin{split}
 \,\lim\limits_{\ell \to 0} \bigg\{ \left(\frac{\ell}{2 \pi}\right)^{2 \Delta_{\cal O}}
\frac{\langle h_1,..,h_n| {\cal O}(0)| h_1,...h_n\rangle}{\langle h_1,...,h_n|h_1,...,h_n\rangle} \bigg\}
=\frac{\langle \Delta|L_{n_1}...L_{n_j}\, {\cal O}(0)\, L_{-n_1}...L_{-n_j}|\Delta \rangle}{\langle \Delta|L_{n_1}...L_{n_j}L_{-n_1}...L_{-n_j}|\Delta \rangle},
\end{split}
\end{equation}
where $\Delta_{\cal O}$ is the scaling dimension of the local operator $\cal O$ in the complex Liouville CFT.  
In order to refer more easily to the soliton states, we introduce the notation:
\begin{equation} \label{notsol}
\begin{split}
|I_1,..,I_n\rangle=|h_1,...,h_n \rangle,
\end{split}
\end{equation}
where $I_j$s are the hole quantum numbers (\ref{Hkvant}) in the NLIE (\ref{ddv1}).
\begin{figure}
\begin{flushleft}
\hskip 25mm
\leavevmode
\epsfxsize=100mm
\epsfbox{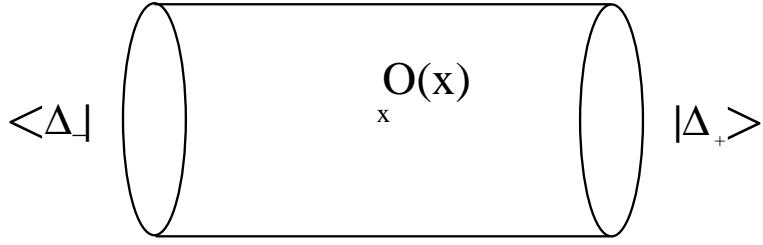}
\end{flushleft}
\caption{\footnotesize
The pictorial representation of the cylindrical geometry for the small volume checks. 
}
\label{YHNLIE}
\end{figure}
In this paper for our numerical checks we use the following pure hole states:
\begin{itemize}
\item $|-\tfrac{1}{2},\tfrac{1}{2}\rangle \to |\Delta\rangle, \qquad \Delta=1-\tfrac{1}{\nu},$
\item $|-\tfrac{1}{2},\tfrac{3}{2}\rangle \to |\Delta,\Delta+1\rangle, \qquad \Delta=1-\tfrac{1}{\nu},$
\item $|-\tfrac{3}{2},\tfrac{3}{2}\rangle \to |\Delta+1,\Delta+1\rangle, \qquad \Delta=1-\tfrac{1}{\nu},$
\item $|-\tfrac{3}{2},-\tfrac{1}{2},\tfrac{1}{2},\tfrac{3}{2}\rangle \to |\Delta,\Delta\rangle, \qquad \Delta=1-\tfrac{2}{\nu},$
\end{itemize}
where on the right hand side the (Liouville) CFT limit of the states are indicated, with the notation $|\Delta+1\rangle=L_{-1} |\Delta \rangle$. 
The set of operators we will consider is as follows:
\begin{equation} \label{opcons}
\begin{split}
{\cal O}(0) \in \{\Phi_{2 \tfrac{1-\nu}{\nu}},\Phi_{4 \tfrac{1-\nu}{\nu}},{\bf l}_{-2} \Phi_{2 \tfrac{1-\nu}{\nu}},{\bf l}_{-2}\cdot 1,,{\bf l}_{-2}^2\cdot 1,{\bf l}_{-4}\cdot 1 \}.
\end{split}
\end{equation}
The Liouville prediction for the case, when both the operator and the sandwiching state associated to some primary field is given by the formula (See section 6. of \cite{SGV}):
\begin{equation} \label{prim2CFT}
\begin{split}
{\cal V}(\alpha,\kappa)\equiv \frac{\langle \Phi_{1-\kappa}(-\infty) \Phi_{\alpha+2\tfrac{1-\nu}{\nu}}(0) \Phi_{1+\kappa}(\infty)\rangle}
{\langle \Phi_{1-\kappa}(-\infty) \Phi_{\alpha}(0) \Phi_{1+\kappa}(\infty) \rangle}=\mu^2 \, \Gamma(\nu)^2 \, Y(x(\alpha)) \, W(\alpha,\kappa) \, W(\alpha,-\kappa),
\end{split}
\end{equation}
where $\mu$ and $x(\alpha)$ are defined in (\ref{muM}) and (\ref{C1}) respectively, and 
\begin{equation} \label{Yx}
\begin{split}
Y(x)&=-2 \, \nu\, x \cdot \frac{\Gamma^2(\nu x+1/2-\nu/2) \, \Gamma(\nu-2 \nu x)}{\Gamma^2(1/2+\nu/2-\nu x) \, \Gamma(2 \nu x+1-\nu)}\cdot 
\frac{\Gamma(-2 \nu x)}{\Gamma(2 \nu x)}, \\
W(\alpha,\kappa)&=\frac{\Gamma(\alpha\nu/2-\nu+1+\kappa \nu)}{\Gamma(-\alpha \nu/2+\nu+\kappa\nu)}.
\end{split}
\end{equation}
Using (\ref{prim2CFT}) it can be shown, that 
\begin{equation} \label{prim4CFT}
\begin{split}
 \frac{\langle \Phi_{1-\kappa}(-\infty) \Phi_{\alpha+4\tfrac{1-\nu}{\nu}}(0) \Phi_{1+\kappa}(\infty)\rangle}
{\langle \Phi_{1-\kappa}(-\infty) \Phi_{\alpha}(0) \Phi_{1+\kappa}(\infty) \rangle}={\cal V}(\alpha,\kappa) \, {\cal V}(\alpha+2\tfrac{1-\nu}{\nu},\kappa).
\end{split}
\end{equation}
With the help of the cylinder conformal Ward-identities, the 3-point functions of descendants can be related to those of the primaries \cite{Boos1,SmirBajn}. 
Again, for short we denote $|\Delta+1 \rangle=L_{-1}|\Delta\rangle.$ Then the following relations are obtained:
\begin{align}
\frac{\langle\mathbf{l}_{-2}\Phi_{\alpha}\rangle_{\Delta}}{\langle \Phi_{\alpha}\rangle_{\Delta}} & =\Delta-\frac{c}{24}-\frac{\Delta_{\alpha}}{12}\,,\label{ratioprim}\\
\frac{\langle\mathbf{l}_{-4}\Phi_{\alpha}\rangle_{\Delta}}{\langle \Phi_{\alpha}\rangle_{\Delta}} & =\frac{\Delta_{\alpha}}{240}\,,\nonumber \\
\frac{\langle\mathbf{l}_{-2}^{2}\Phi_{\alpha}\rangle_{\Delta}}{\langle \Phi_{\alpha}\rangle_{\Delta}} & =\Delta^{2}-\Delta\frac{2\Delta_{\alpha}+c+2}{12}+\frac{20\Delta_{\alpha}^{2}+56\Delta_{\alpha}+20c\Delta_{\alpha}+5c^{2}+22c}{2880}\,.\nonumber 
\end{align}

\begin{align}
\frac{\langle \Phi_{\alpha}\rangle_{\Delta+1}}{\langle \Phi_{\alpha}\rangle_{\Delta}}\quad & =2\Delta+\Delta_{\alpha}^{2}-\Delta_{\alpha}\,,\label{ratiodesc}\\
\frac{\langle\mathbf{l}_{-2}\Phi_{\alpha}\rangle_{\Delta+1}}{\langle \Phi_{\alpha}\rangle_{\Delta}} & =2\Delta^{2}+\Delta\frac{12\Delta_{\alpha}^{2}+34\Delta_{\alpha}+24-c}{12}-\frac{(\Delta_{\alpha}-1)\Delta_{\alpha}(2\Delta_{\alpha}-24+c)}{24}\,,\nonumber \\
\frac{\langle\mathbf{l}_{-4}\Phi_{\alpha}\rangle_{\Delta+1}}{\langle \Phi_{\alpha}\rangle_{\Delta}} & =\Delta\frac{241\Delta_{\alpha}}{120}-\frac{\Delta_{\alpha}^{2}}{240}+\frac{\Delta_{\alpha}^{3}}{240}\,,\nonumber \\
\frac{\langle\mathbf{l}_{-2}^{2}\Phi_{\alpha}\rangle_{\Delta+1}}{\langle \Phi_{\alpha}\rangle_{\Delta}} & =2\Delta^{3}+\Delta^{2}\frac{70-c+40\Delta_{\alpha}+6\Delta_{\alpha}^{2}}{6}\nonumber \\
 & \,\,\,+\Delta\frac{2400-218c+5c^{2}+4616\Delta_{\alpha}-340c\Delta_{\alpha}+1940\Delta_{\alpha}^{2}-120c\Delta_{\alpha}^{2}-240\Delta_{\alpha}^{3}}{1440}\nonumber \\
 & \,\,\,+\frac{(\Delta_{\alpha}-1)\Delta_{\alpha}(2400-218c+5c^{2}-424\Delta_{\alpha}+20c\Delta_{\alpha}+20\Delta_{\alpha}^{2})}{2880}\,\,.\nonumber 
\end{align}

Now, we are in the position to check whether the ultraviolet limit our formulas 
(\ref{Gj}), (\ref{omkif}) and (\ref{expform}) 
for the expectation values is described by the complex Liouville CFT.  
Unfortunately, we can not solve analytically the equations in the small volume limit, thus 
the comparison with CFT predictions, is performed through the numerical solution of the 
equations.  
 In the rest of this section,
we set ${\cal M}=1$ and measure everything in the units of soliton-mass.

\subsection{The case of primaries $\Phi_{2\tfrac{1-\nu}{\nu}}(0)$ and $\Phi_{4\tfrac{1-\nu}{\nu}}(0)$}

The expectation values of these operators are given by the formulas (\ref{expform}):
\begin{equation} \label{Fi24}
\begin{split}
\langle \Phi_{2\tfrac{1-\nu}{\nu}}(0)\rangle&=C_1(0) \, \Pi(\nu)^{2-2\nu} \, \left( \omega_{1,-1}(0)+\tfrac{i}{\nu} \cot\tfrac{\pi}{2 \nu}\right), \\
\langle \Phi_{4\tfrac{1-\nu}{\nu}}(0)\rangle&=C_2(0) \, \Pi(\nu)^{8(1-\nu)} \, 
\left\{ -\tfrac{1}{\nu^2} \cot \tfrac{\pi}{2 \nu}\, \cot \tfrac{3 \pi}{2 \nu}+
\tfrac{i}{\nu} \cot\tfrac{\pi}{2 \nu} \omega_{3,-3}(0)+\tfrac{i}{\nu} \cot\tfrac{3\pi}{2 \nu} \omega_{1,-1}(0)+\right.\\
 &\omega_{1,-1}(0)\, \omega_{3,-3}(0)-\omega_{1,-3}(0) \omega_{3,-1}(0)   \big\}.
\end{split}
\end{equation}
Let us define the CFT limit of a finite volume expectation value by the formula:
\begin{equation} \label{CFTlimdef}
\begin{split}
\langle {\cal O}(0) \rangle_{CFT}=\lim\limits_{\ell \to 0} \left\{ \langle {\cal O}(0)\rangle \left(\frac{\ell}{2 \pi}\right)^{2 \Delta_{\cal O}}\right\}.
\end{split}
\end{equation}
Then for the four states we considered under (\ref{notsol}), the CFT computations imply the following small volume limits for the primaries $\Phi_{2m(1-\nu)/\nu}$: 
\begin{itemize}
\item $|-\tfrac{1}{2},\tfrac{1}{2}\rangle:$
\begin{equation} \label{ppCFT2}
\begin{split}
\langle \Phi_{2 m\tfrac{1-\nu}{\nu}}\rangle_{CFT}=\langle \Phi_{2 m\tfrac{1-\nu}{\nu}}\rangle_{\Delta=1-\tfrac{1}{\nu}},
\end{split}
\end{equation}
\item $|-\tfrac{3}{2},-\tfrac{1}{2},\tfrac{1}{2},\tfrac{3}{2}\rangle:$
\begin{equation} \label{ppCFT4}
\begin{split}
\langle \Phi_{2 m\tfrac{1-\nu}{\nu}}\rangle_{CFT}=\langle \Phi_{2 m\tfrac{1-\nu}{\nu}}\rangle_{\Delta=1-\tfrac{2}{\nu}},
\end{split}
\end{equation}
\item $|-\tfrac{1}{2},\tfrac{3}{2}\rangle:$
\begin{equation} \label{ppCFT2d1}
\begin{split}
\langle \Phi_{2 m\tfrac{1-\nu}{\nu}}\rangle_{CFT}
=  \tfrac{2 \Delta+\Delta_{2m(1-\nu)/\nu}^2-\Delta_{2m(1-\nu)/\nu}}{2 \Delta}  \langle \Phi_{2 m\tfrac{1-\nu}{\nu}}\rangle\bigg|_{\Delta=1-\tfrac{1}{\nu}},
\end{split}
\end{equation}
\item $|-\tfrac{3}{2},\tfrac{3}{2}\rangle:$
\begin{equation} \label{ppCFT2dd1}
\begin{split}
\langle \Phi_{2 m\tfrac{1-\nu}{\nu}}\rangle_{CFT}
= \left( \tfrac{(2 \Delta+\Delta_{2m(1-\nu)/\nu}^2-\Delta_{2m(1-\nu)/\nu})^2}{2 \Delta} \right)^2
 \langle \Phi_{2 m\tfrac{1-\nu}{\nu}}\rangle\bigg|_{\Delta=1-\tfrac{1}{\nu}},
\end{split}
\end{equation}
\end{itemize}
where according to (\ref{prim2CFT}):
\begin{equation} \label{CFTprimlim}
\begin{split}
\langle \Phi_{2 m\tfrac{1-\nu}{\nu}}\rangle_{\Delta=1-\kappa}=\prod\limits_{j=0}^{m-1}{\cal V}(2 j \tfrac{1-\nu}{\nu},\kappa).
\end{split}
\end{equation}
The numerical results are obtained at $\nu=\tfrac{4}{7}$ are summarized in table 1. The CFT limit of the numerical results are 
taken from computing $\langle {\cal O}(0)\rangle \left(\frac{\ell}{2 \pi}\right)^{2 \Delta_{\cal O}}$
at $\ell=10^{-2}.$

\begin{table}[h]
\begin{center}
\begin{tabular}{|c|c|c|c|c|}
\hline
 & \multicolumn{2}{c|}{$\langle \Phi_{2 \tfrac{1-\nu}{\nu}}\rangle_{CFT}$}   & \multicolumn{2}{c|}{$\langle \Phi_{4 \tfrac{1-\nu}{\nu}}\rangle_{CFT}$}  \tabularnewline
\hline 
 state  & \text{Numerics} & \text{CFT prediction}  & \text{Numerics} & \text{CFT prediction}  \\
 \hline
$|-\tfrac{1}{2},\tfrac{1}{2}\rangle$ & 2.7862399017 & 2.7862399479 & 0.7017902707  & 0.7017903948  \\
\hline
$|-\tfrac{3}{2},-\tfrac{1}{2},\tfrac{1}{2},\tfrac{3}{2}\rangle$ & 2.3026776339 & 2.3026776429 & 1.5315414850 & 1.5315415383 \\
\hline
$|-\tfrac{1}{2},\tfrac{3}{2}\rangle$ & 3.3651982417 & 3.3651988981 & 0.4830504382 & 0.4830505315 \\
\hline
$|-\tfrac{3}{2},\tfrac{3}{2}\rangle$ & 4.0644590918 & 4.0644610068 & 0.3324887760 & 0.3324893268 \\
\hline
\end{tabular}\label{tablepro1}
\bigskip
\caption{Numerical data for the expectation values of primaries $\Phi_{2m (1-\nu)\nu}(0),$ with $m=1,2,$ at the $(\nu=\tfrac{4}{7}, \ell=10^{-2})$ point of the parameter space.}
\label{text1}
\end{center}
\end{table}
\normalsize

\subsection{Expectation values of descendant fields}

In order to consider expectation values of descendant fields, one needs to know how the fermionic basis in (\ref{basis}) is related to the 
Virasoro basis used in the Liouville 3-point functions. For our studies we need to know this relation upto some low lying levels. The relation is 
known upto level 8 \cite{Boos1}. Nevertheless,  
here we summarize from \cite{SGV}, only the most important formulas, we need. 

The fermionic basis can be expressed in terms of the Virasoro descendants according to formula:
\begin{equation} \label{fermvir}
\begin{split}
\beta_{I^+}^* \gamma_{I^-}^* \Phi_\alpha(0)=\prod\limits_{2j-1 \in I^+} D_{2j-1}(\alpha) \prod\limits_{2j-1 \in I^-} D_{2j-1}(2-\alpha) 
[P^{even}_{I^+,I^-}+d_\alpha P^{odd}_{I^+,I^-}],
\end{split}
\end{equation}
where 
\begin{equation} \label{dalf}
\begin{split}
&d_\alpha=\frac{\nu(\nu-2)}{\nu-1} (\alpha-1), \\
&D_{2j-1}(\alpha)=-\sqrt{\frac{i}{\nu}} \, \Gamma(\nu)^{-\tfrac{2j-1}{\nu}} \, (1-\nu)^{\tfrac{2j-1}{2}} \frac{\Gamma\left(\tfrac{\alpha}{2}+\tfrac{1}{2 \nu}(2j-1)\right)}
{(j-1)! \, \Gamma\left( \tfrac{\alpha}{2}+\tfrac{1-\nu}{2 \nu}(2j-1) \right)},
\end{split}
\end{equation}
and $P^{even}_{I^+,I^-}$ and $P^{odd}_{I^+,I^-}$ are polynomial expressions of the even Virasoro generators $\{ {\bf l}_{-2k}\}.$ For the first few levels they take the form as follows \cite{SGV}:  
\begin{equation} \label{Pk}
\begin{split}
&P^{even}_{\{1\},\{1\}}={\bf l}_{-2}, \qquad P^{odd}_{\{1\},\{1\}}=0, \\
&P^{even}_{\{1\},\{3\}}=P^{even}_{\{3\},\{1\}}=\tfrac{1}{2}{\bf l}_{-2}^2+\tfrac{c-16}{9}{\bf l}_{-4}, \qquad P^{odd}_{\{1\},\{3\}}=P^{odd}_{\{3\},\{1\}}=-\tfrac{1}{3} {\bf l}_{-4}.
\end{split}
\end{equation}
The  descendants of the "shifted-primaries" $\Phi_{\alpha+2n\tfrac{1-\nu}{\nu}}(0)$ can be computed from the formula (\ref{mexpform}).

\begin{table}[h]
\begin{center}
\begin{tabular}{|c|c|c|c|c|}
\hline
& \multicolumn{2}{c|}{$\langle {\bf l}_{-2} \Phi_{2 \tfrac{1-\nu}{\nu}}\rangle_{CFT}$} & \multicolumn{2}{c|}{$\omega_{1,1}(0)_{CFT}$}     \tabularnewline
\hline 
 state  & \text{Numerics} & \text{CFT prediction}  & \text{Numerics} & \text{CFT prediction}  \\
 \hline
$|-\tfrac{1}{2},\tfrac{1}{2}\rangle$ &  1.5423829631 & 1.5423828283  & 1.2069251545 & 1.2069249851 \\
\hline
$|-\tfrac{3}{2},-\tfrac{1}{2},\tfrac{1}{2},\tfrac{3}{2}\rangle$ & 5.3043825168 & 5.3043824274 & 5.1062212309 i & 5.1062210908 i \\
\hline
$|-\tfrac{1}{2},\tfrac{3}{2}\rangle$ & 4.4320078359 & 4.432008303 & 3.4350943930 i & 3.4350941884 i \\
\hline
$|-\tfrac{3}{2},\tfrac{3}{2}\rangle$ & 5.3529430027 & 5.3529450933 & 3.4350944355 i &  3.4350941884 i \\
\hline
\end{tabular}\label{tablepro2}
\bigskip
\caption{Numerical data for the expectation values of the level 2 descendant fields; 
${\bf l}_{-2} \Phi_{2 (1-\nu)/\nu}$ and $ {\bf l}_{-2} 1,$  at the $(\nu=\tfrac{4}{7}, \ell=10^{-2})$ point of the parameter space.}
\label{text2}
\end{center}
\end{table}
\normalsize

\subsubsection{The case of $\langle {\bf l}_{-2}\, \Phi_{2 \tfrac{1-\nu}{\nu}}\rangle$}

Using (\ref{mexpform}), (\ref{fermvir}) and (\ref{fermant}) one obtains the following expression for $\langle {\bf l}_{-2} \Phi_{2 \tfrac{1-\nu}{\nu}}\rangle:$
\begin{equation} \label{lm2PHI}
\begin{split}
\langle {\bf l}_{-2} \Phi_{2 \tfrac{1-\nu}{\nu}}\rangle=\frac{C_1(0) \, \tfrac{1}{i \, \nu}\, \cot \tfrac{\pi}{2 \nu} \, \Pi(\nu)^{4-2\nu}  }{D_1\left(2 \tfrac{1-\nu}{\nu}\right) \,
D_1\left(2-2 \tfrac{1-\nu}{\nu}\right)} \, \omega_{3,-1}(0).
\end{split}
\end{equation}
Then for the four states under consideration the Liouville CFT gives the following predictions: 
\begin{itemize}
\item $|-\tfrac{1}{2},\tfrac{1}{2}\rangle:$
\begin{equation} \label{1ppCFT2}
\begin{split}
\langle  {\bf l}_{-2} \Phi_{2 \tfrac{1-\nu}{\nu}}\rangle_{CFT}=\left( \Delta-\frac{c}{24}-\frac{\Delta_{2(1-\nu)/\nu}}{12}\right)\, \langle \Phi_{2 \tfrac{1-\nu}{\nu}}\rangle\bigg|_{\Delta=1-\tfrac{1}{\nu}},
\end{split}
\end{equation}
\item $|-\tfrac{3}{2},-\tfrac{1}{2},\tfrac{1}{2},\tfrac{3}{2}\rangle:$
\begin{equation} \label{1ppCFT4}
\begin{split}
\langle  {\bf l}_{-2} \Phi_{2 \tfrac{1-\nu}{\nu}}\rangle_{CFT}= \left( \Delta-\frac{c}{24}-\frac{\Delta_{2(1-\nu)/\nu}}{12}\right)\, \langle \Phi_{2 \tfrac{1-\nu}{\nu}}\rangle\bigg|_{\Delta=1-\tfrac{2}{\nu}},
\end{split}
\end{equation}
\item $|-\tfrac{1}{2},\tfrac{3}{2}\rangle:$
\begin{equation} \label{1ppCFT2d1}
\begin{split}
\langle  {\bf l}_{-2} \Phi_{2 \tfrac{1-\nu}{\nu}}\rangle_{CFT}
=  \frac{F_{-2}(\Delta,\Delta_{2 (1-\nu)/\nu})}{2 \Delta}  \langle \Phi_{2 \tfrac{1-\nu}{\nu}}\rangle\bigg|_{\Delta=1-\tfrac{1}{\nu}},
\end{split}
\end{equation}
\item $|-\tfrac{3}{2},\tfrac{3}{2}\rangle:$
\begin{equation} \label{1ppCFT2dd1}
\begin{split}
\langle  {\bf l}_{-2} \Phi_{2 \tfrac{1-\nu}{\nu}}\rangle_{CFT}
=  \tfrac{2 \Delta+\Delta_{2(1-\nu)/\nu}^2-\Delta_{2(1-\nu)/\nu}}{(2 \Delta)^2} \, F_{-2}(\Delta,\Delta_{2 (1-\nu)/\nu}) \, \langle \Phi_{2 \tfrac{1-\nu}{\nu}}\rangle\bigg|_{\Delta=1-\tfrac{1}{\nu}},
\end{split}
\end{equation}
\end{itemize}
where we introduced the function from the right hand side of second line of (\ref{ratiodesc}):
\begin{equation} \label{Fm2}
\begin{split}
F_{-2}(\Delta,\Delta_\alpha)=2\Delta^{2}+\Delta\frac{12\Delta_{\alpha}^{2}+34\Delta_{\alpha}+24-c}{12}-\frac{(\Delta_{\alpha}-1)\Delta_{\alpha}(2\Delta_{\alpha}-24+c)}{24}.
\end{split}
\end{equation}
The corresponding numerical results obtained at $\ell=10^{-2}$ and $\nu=\tfrac{4}{7},$ 
can be found in table 2.
\subsubsection{Expectation values of the descendants of the unity}

Now, instead of expressing the expectation values of the operators under consideration, we express the $\omega_{ij}(\alpha)$ matrix elements in terms of the 
expectation values of the low lying descendants of the primary field $\Phi_\alpha(0).$
The formulas one obtains are as follows: 
\begin{equation} \label{omegasl}
\begin{split}
\omega_{1,1}(\alpha)&=\mu^{-2/\nu} \, D_1(\alpha) \, D_1(2-\alpha) \, \frac{\langle{\bf l}_{-2} \Phi_\alpha(0)\rangle}{\langle \Phi_\alpha(0)\rangle}, \\
\omega_{3,1}(\alpha)&=\mu^{-4/\nu} \, D_3(\alpha) \, D_1(2-\alpha) \,\left\{ \frac{1}{2} \frac{\langle{\bf l}_{-2}^2 \Phi_\alpha(0)\rangle}{\langle \Phi_\alpha(0)\rangle}+
\left(\frac{c-16}{9}+\frac{d_\alpha}{3} \right)\frac{\langle{\bf l}_{-4} \Phi_\alpha(0)\rangle}{\langle \Phi_\alpha(0)\rangle} \right\},\\
\omega_{1,3}(\alpha)&=\mu^{-4/\nu} \, D_1(\alpha) \, D_3(2-\alpha) \,\left\{ \frac{1}{2} \frac{\langle{\bf l}_{-2}^2 \Phi_\alpha(0)\rangle}{\langle \Phi_\alpha(0)\rangle}+
\left(\frac{c-16}{9}-\frac{d_\alpha}{3} \right)\frac{\langle{\bf l}_{-4} \Phi_\alpha(0)\rangle}{\langle \Phi_\alpha(0)\rangle} \right\}.
\end{split}
\end{equation}
For the comparison of numerical and the CFT results, we define the CFT limit of the above $\omega$s as follows: 
\begin{equation} \label{omCFT}
\begin{split}
\omega_{jk}(\alpha)_{CFT}=\lim\limits_{\ell \to 0} \left\{ \omega_{jk}(\alpha) \, \left(\tfrac{\ell}{2 \pi}\right)^{\text{dim} \, \omega_{jk}(\alpha)}  \right\},
\end{split}
\end{equation}
where $\text{dim} \omega_{jk}(\alpha)$ is the length dimension of $\omega_{jk}(\alpha).$ For the 3 special cases in (\ref{omegasl}) 
$\text{dim} \, \omega_{jk}(\alpha)=j+k.$
We check (\ref{omegasl}) in the $\alpha \to 0$ limit. As a consequence of (\ref{ratioprim}) and (\ref{ratiodesc}), 
in this case in the small volume limit $\langle {\bf l}_{-4} \, 1\rangle=0.$ Thus,  $\omega_{1,3}(0)_{CFT}=\omega_{3,1}(0)_{CFT}.$ 

Using the formulas (\ref{ratioprim}) and (\ref{ratiodesc}), for the four specific states, 
one obtains the CFT predictions as follows:
\begin{itemize}
\item $|-\tfrac{1}{2},\tfrac{1}{2}\rangle:$
\begin{equation} \label{2ppCFT2}
\begin{split}
\omega_{1,1}(0)_{CFT}=\Pi(\nu)^{-2} \, D_1(0) \, D_1(2) \, 
\left( \Delta-\frac{c}{24}\right)\, \bigg|_{\Delta=1-\tfrac{1}{\nu}},
\end{split}
\end{equation}
\item $|-\tfrac{3}{2},-\tfrac{1}{2},\tfrac{1}{2},\tfrac{3}{2}\rangle:$
\begin{equation} \label{2ppCFT4}
\begin{split}
\omega_{1,1}(0)_{CFT}=\Pi(\nu)^{-2} \, D_1(0) \, D_1(2) \, 
\left( \Delta-\frac{c}{24}\right)\, \bigg|_{\Delta=1-\tfrac{2}{\nu}},
\end{split}
\end{equation}
\item $|-\tfrac{1}{2},\tfrac{3}{2}\rangle:$
\begin{equation} \label{2ppCFT2d1}
\begin{split}
\omega_{1,1}(0)_{CFT}
= \Pi(\nu)^{-2} \, D_1(0) \, D_1(2) \, \frac{F_{-2}(\Delta,0)}{2 \Delta}  \bigg|_{\Delta=1-\tfrac{1}{\nu}},
\end{split}
\end{equation}
\item $|-\tfrac{3}{2},\tfrac{3}{2}\rangle:$
\begin{equation} \label{2ppCFT2dd1}
\begin{split}
\omega_{1,1}(0)_{CFT}
= \Pi(\nu)^{-2} \, D_1(0) \, D_1(2) \, \frac{F_{-2}(\Delta,0)}{2 \Delta}  \bigg|_{\Delta=1-\tfrac{1}{\nu}}.
\end{split}
\end{equation}
\end{itemize}
Before turning to the case of $\omega_{3,1}(0)_{CFT},$ from (\ref{ratioprim}) and (\ref{ratiodesc}) taken at $\Delta_\alpha=0$, it is worth to introduce the 
functions as follows:
\begin{equation} \label{lm22Phi}
\begin{split}
F_0(\Delta)&=\langle {\bf l}_{-2}^2 \cdot 1\rangle_\Delta=\Delta^2-\Delta\, \frac{c+2}{12}+\frac{5 c^2+22\, c}{2880}, \\
F_1(\Delta)&=\frac{\langle {\bf l}_{-2}^2 \cdot 1\rangle_{\Delta+1}}{\langle 1 \rangle_\Delta}
=2 \Delta^3+\Delta^2\,\frac{70-c}{6}+
\Delta \,\frac{5 c^2-218 c+2400}{1440}.
\end{split}
\end{equation}
Then one obtains the following results in the CFT limit:
\begin{itemize}
\item $|-\tfrac{1}{2},\tfrac{1}{2}\rangle:$
\begin{equation} \label{32ppCFT2}
\begin{split}
\omega_{3,1}(0)_{CFT}=\Pi(\nu)^{-4} \, D_3(0) \, D_1(2) \, \frac{1}{2}
F_0\left( \Delta\right)\, \bigg|_{\Delta=1-\tfrac{1}{\nu}},
\end{split}
\end{equation}
\item $|-\tfrac{3}{2},-\tfrac{1}{2},\tfrac{1}{2},\tfrac{3}{2}\rangle:$
\begin{equation} \label{32ppCFT4}
\begin{split}
\omega_{3,1}(0)_{CFT}=\Pi(\nu)^{-4} \, D_3(0) \, D_1(2) \, \frac12
F_0\left( \Delta\right)\, \bigg|_{\Delta=1-\tfrac{2}{\nu}},
\end{split}
\end{equation}
\item $|-\tfrac{1}{2},\tfrac{3}{2}\rangle:$
\begin{equation} \label{32ppCFT2d1}
\begin{split}
\omega_{3,1}(0)_{CFT}
= \Pi(\nu)^{-4} \, D_3(0) \, D_1(2) \, \frac{F_1(\Delta)}{4 \Delta} \bigg|_{\Delta=1-\tfrac{1}{\nu}},
\end{split}
\end{equation}
\item $|-\tfrac{3}{2},\tfrac{3}{2}\rangle:$
\begin{equation} \label{32ppCFT2dd1}
\begin{split}
\omega_{3,1}(0)_{CFT}
= \Pi(\nu)^{-4} \, D_3(0) \, D_1(2) \, \frac{F_1(\Delta)}{4 \Delta} \bigg|_{\Delta=1-\tfrac{1}{\nu}}.
\end{split}
\end{equation}
\end{itemize}
The corresponding numerical results obtained at $\ell=10^{-2}$ and $\nu=\tfrac{4}{7},$ 
can be found in tables 2. and 3.

\begin{table}[h]
\begin{center}
\begin{tabular}{|c|c|c|c|c|}
\hline
 & \multicolumn{1}{c}{$\omega_{1,3}(0)_{CFT}$} &  & \multicolumn{1}{c}{$\omega_{3,1}(0)_{CFT}$} & \tabularnewline
\hline 
 state  & \text{Numerics} & \text{CFT prediction}  & \text{Numerics} & \text{CFT prediction}  \\
 \hline
$|-\tfrac{1}{2},\tfrac{1}{2}\rangle$  & 1.9427100198 i  & 1.9427092046 i & 1.9427100198 i & 1.9427092046 i  \\
\hline
$|-\tfrac{3}{2},-\tfrac{1}{2},\tfrac{1}{2},\tfrac{3}{2}\rangle$ & 47.1269102712 i & 47.1269074676 i &  47.1269102712 i & 47.1269074676 i \\
\hline
$|-\tfrac{1}{2},\tfrac{3}{2}\rangle$ & 35.7155944324 i & 35.7155920899 i &  35.7155944324 i & 35.7155920899 i  \\
\hline
$|-\tfrac{3}{2},\tfrac{3}{2}\rangle$ & 35.7155949192 i & 35.7155920899 i &  35.7155949192 i & 35.7155920899 i \\
\hline
\end{tabular}\label{tableproo13}
\bigskip
\caption{Numerical data for the expectation values of the level 4 descendants of the unity,  
  at the $(\nu=\tfrac{4}{7}, \ell=10^{-2})$ point of the parameter space.}
\label{text3}
\end{center}
\end{table}
\normalsize



\section{Summary and Conclusions} \label{concl}

In the fundamental paper \cite{SGV}, 1-point functions or in other words ground state expectation values of exponential fields and 
of their Virasoro descendants have been determined in the sine-Gordon model defined in cylindrical geometry. 
The derivation was done in the so-called fermionic basis discovered in the integrable  light-cone lattice regularization 
\cite{ddvlc} of the model{\footnote{Namely, in the 6-vertex model with appropriate alternating inhomogeneities.}}.
The relation of the fermionic basis of operators to that of the Virasoro basis, has been determined upto level 8 \cite{Boos1}. 
The main advantage of the fermionic basis is that in this basis, the 1-point functions can be given by simple determinant 
expressions of a single nontrivial matrix $\omega_{jk}(\alpha),$ whose matrix elements can be obtained by solving linear 
integral equations containing the counting-function of the ground state as a fundamental building block of its kernel. 

In this paper, we extended these results  from ground state expectation values to expectation values in any eigenstate of the 
Hamiltonian of the sine-Gordon model. Structurally all formulas of \cite{SGV} remain the same, but in the case of expectation values
in excited states the matrix $\omega_{jk}(\alpha)$ will depend on the sandwiching state. 
Thus, the linear equations, which determine the matrix $\omega_{jk}(\alpha),$ are needed to be modified by appropriate source terms, 
which characterize the sandwiching state.

Having these equations at hand, we performed several tests on the equations in the pure multi-soliton sector. 
In the large volume limit we performed three tests. We have shown, that the excited state formulation of the matrix $\omega_{jk}(\alpha)$ 
still respects the nonlinear compatibility equation, which is necessary to get unique answer for each expectation value of each operator. 
We also checked, that our excited state formulation of the expectation values is consistent with the previous LeClair-Mussardo type 
large volume series conjecture of \cite{En}. Finally, from the multi-soliton expectation values of 
the set of primaries $\Phi_{2 \, n(1-\nu)/\nu}, \quad n\in \mathbb{N},$ we computed the classical limit of connected diagonal form-factors upto two soliton states, 
and compared them to the results of semi-classical computations coming from direct field theoretical computations of ref. \cite{BJsg}. The comparison gave perfect agreement.

The equations were checked in the ultraviolet limit, as well. 
In this limit, the expectation values tend to 3-point functions of the complex Liouville conformal field theory. 
While, in  the large volume limit the equations determining the expectation values, can be solved analytically, 
in the ultraviolet limit they can be solved only numerically. 
Nevertheless, the accurate solution of the equations allowed us to check our results for various operators 
and low lying multi-soliton excitations, 
against Liouville  3-point functions.  
Perfect agreement was found in this regime, too. 

The sine-Gordon was the first integrable model with non-diagonally scattering theory, where all finite volume diagonal 
matrix elements of local operators could have been determined. The volume dependence in the final formulas 
for the expectation values is encoded into the volume dependence of the counting-function. This function satisfies 
certain nonlinear integral equations and governs the finite volume dependence of the energy of the sandwiching state. 
This description of the finite volume spectrum is very specific for the sine-Gordon model and there is no general method to 
find an analogue of this description to other models. For some specific examples see references \cite{BHhnlie,SSh,KLsun,GKVsu2} etc. 
Nevertheless, there is another method to describe the finite volume spectrum in integrable quantum field theories 
with non-diagonally scattering theory. This is Thermodynamic Bethe Ansatz (TBA) method, where the rapidity dependent 
pseudo-energies satisfy the nonlinear TBA integral equations and govern the volume dependence of the energies.  
Though the concrete form of the TBA equations are more complicated than that of the counting function, this formulation 
has the advantage, that it can be generalized to any non-diagonally scattering theory. Thus, rephrasing the results of this paper 
in the language of the TBA pseudo-energies, could give a deeper insight into the structure of the formulas and help 
in finding a generalization to other important models.



\vspace{1cm}
{\tt Acknowledgments}

\noindent 
The author would like to thank Zolt\'an Bajnok and Fedor Smirnov for useful discussions.
This work was supported by the NKFIH research Grant K116505
and by a Hungarian-French bilateral exchange project.

\appendix








\newpage

\begin{thebibliography}{99}



\bibitem{BJsftv}
  Z.~Bajnok and R.~A.~Janik,
  ``String field theory vertex from integrability,''
  {\em JHEP  1504} (2015) 042,
  [arXiv:1501.04533 [hep-th]].


\bibitem{BJhhl}
Z.~{Bajnok}, R.~A. {Janik}, and A.~{Wereszczynski}, ``{HHL correlators, orbit
  averaging and form factors},''
{{\em Journal of High Energy Physics} {\bf 9} (Sept., 2014) 50},
  [arXiv:1404.4556 [hep-th]].
\bibitem{Hollo}
 L.~Hollo, Y.~Jiang and A.~Petrovskii,
  ``Diagonal Form Factors and Heavy-Heavy-Light Three-Point Functions at Weak Coupling,''
  JHEP {\bf 1509} (2015) 125
  [arXiv:1504.07133 [hep-th]].
  %
  \bibitem{JF1}
  Y.~Jiang and A.~Petrovskii,
  ``Diagonal form factors and hexagon form factors,''
  JHEP {\bf 1607} (2016) 120,
  [arXiv:1511.06199 [hep-th]].
  %
  \bibitem{JF2}
  Y.~Jiang,
  ``Diagonal Form Factors and Hexagon Form Factors II. Non-BPS Light Operator,''
  JHEP {\bf 1701} (2017) 021
  [arXiv:1601.06926 [hep-th]].
\bibitem{KoEss}
  F.~H.~L.~Essler and R.~M.~Konik,
  ``Applications of massive integrable quantum field theories to problems in condensed matter physics,''
  In *Shifman, M. (ed.) et al.: From fields to strings, vol. 1* 684-830,
  [cond-mat/0412421 [cond-mat.str-el]].
 
\bibitem{PT08a}
B.~Pozsgay and G.~Tak\'acs, ``{Form-factors in finite volume I: Form-factor
  bootstrap and truncated conformal space},''
{{\em Nucl.Phys.} {\bf B788} (2008) 167--208}, [arXiv:0706.1445 [hep-th]].
\bibitem{PT08b}
B.~Pozsgay and G.~Tak\'acs, ``{Form factors in finite volume. II. Disconnected
  terms and finite temperature correlators},''
{{\em Nucl.Phys.} {\bfseries B788} (2008) 209--251},
 [arXiv:0706.3605 [hep-th]].

\bibitem{Pmu}
B.~Pozsgay, ``{Luscher's mu-term and finite volume bootstrap principle for
  scattering states and form factors},''
  {{\em Nucl.Phys.}
  {\bfseries B802} (2008) 435--457},
[arXiv:0803.4445 [hep-th]]  
 
\bibitem{BBCL}
Z.~Bajnok, J.~Balog, M.~Lájer and C.~Wu,
  ``Field theoretical derivation of Lüscher’s formula and calculation of finite volume form factors,''
  JHEP {\bf 1807} (2018) 174, 
  [arXiv:1802.04021 [hep-th]].
  
 \bibitem{BBCL1}
  Z.~Bajnok, M.~Lajer, B.~Szepfalvi and I.~Vona,
  ``Leading exponential finite size corrections for non-diagonal form factors,''
  JHEP {\bf 1907} (2019) 173
  doi:10.1007/JHEP07(2019)173
  [arXiv:1904.00492 [hep-th]]. 


\bibitem{CFT4}
  H.~Boos, M.~Jimbo, T.~Miwa and F.~Smirnov,
  Commun.\ Math.\ Phys.\  {\bf 299} (2010) 825
  doi:10.1007/s00220-010-1051-6
  [arXiv:0911.3731 [hep-th]].  
  
\bibitem{SGV}
  M.~Jimbo, T.~Miwa and F.~Smirnov,
  ``Hidden Grassmann structure in the XXZ model V: Sine-Gordon model,''
  Lett.\ Math.\ Phys.\  {\bf 96} (2011) 325,
  [arXiv:1007.0556 [hep-th]].  
  
\bibitem{SmirBab}
  C.~Babenko and F.~Smirnov,
  ``One point functions of fermionic operators in the Super Sine Gordon model,''
  arXiv:1905.09602 [hep-th].  
  
\bibitem{SmirNeg}
  S.~Negro and F.~Smirnov,
  ``On one-point functions for sinh-Gordon model at finite temperature,''
  Nucl.\ Phys.\ B {\bf 875} (2013) 166,
  [arXiv:1306.1476 [hep-th]].
  
\bibitem{SmirBajn}
  Z.~Bajnok and F.~Smirnov,
  ``Diagonal finite volume matrix elements in the sinh-Gordon model,''
  Nucl.\ Phys.\ B {\bf } (2019) 114664,
  [arXiv:1903.06990 [hep-th]].
  
  


\bibitem{BWu}
  Z.~Bajnok and C.~Wu,
  ``Diagonal form factors from non-diagonal ones,''
  {\em 2017 MATRIX annals 141-151.}
  arXiv:1707.08027 [hep-th].


\bibitem{Palmai13}
T.~P\'almai and G.~Tak\'acs, ``{Diagonal multisoliton matrix elements in finite
  volume},'' {{\em
  Phys.Rev.} {\bf D87} no.~4, (2013) 045010}, [arXiv:1209.6034 [hep-th]]. 

\bibitem{LM99}
A.~Leclair and G.~Mussardo, ``{Finite temperature correlation functions in
  integrable QFT},''
  {\em Nucl.Phys.}{\bf B552} (1999) 624--642, 
[arXiv:9902075 [hep-th]].


\bibitem{Pozsg13}
B.~Pozsgay, ``{Form factor approach to diagonal finite volume matrix elements
  in Integrable QFT},''
{\em  JHEP} {\bf 1307} (2013) 157,
 [arXiv:1305.3373 [hep-th]]. 
  
\bibitem{PST14}
B. Pozsgay, I.~Sz\'ecs\'enyi and G.~Tak\'acs, ``{Exact finite volume expectation values of local operators 
in excited states},''
{\em JHEP 1504} (2015) 023,
  , [arXiv:1412.8436 [hep-th]].  



  
\bibitem{saleur}  
H. Saleur,
 ``{A comment on finite temperature correlations in integrable QFT},''
{\em Nucl.Phys.}{\bf B567} (2000) 602-610,
 [arXiv:hep-th/9909019].  
  
\bibitem{Pozsg11}
B.~Pozsgay, ``{Mean values of local operators in highly excited Bethe
  states},''
 {{\em
  J.Stat.Mech.} {\bf 1101} (2011) P01011},
 [arXiv:1009.4662 [hep-th]].  
  

\bibitem{En}
\'A. Heged\H{u}s,
  ``Lattice approach to the finite volume form-factors of the Massive Thirring
 (sine-Gordon) model,''
  JHEP {\bf 08} (2017) 059,
  [arXiv:1705.0039 [hep-th]].
  
\bibitem{En1}
\'A. Heged\H{u}s,
  ``Exact finite volume expectation values of $\bar{\Psi}\Psi$ in the Massive Thirring model from light-cone lattice correlators ,''
  JHEP {\bf } (2018) ,
  [arXiv:1710.09583 [hep-th]].
%



\bibitem{ddvlc} C. Destri and H.J. de Vega, ``Light-cone lattice approach to fermionic theories in 2D,''
\emph{Nucl. Phys.} \textbf{B290} (1987) 363-391.



\bibitem{KP1}
A. Kl\"umper,M. Batchelor, P. Pearce,
``Central charges of the 6- and 19-vertex models with twisted boundary conditions,''
{\em J. Phys. A.}{\bf 24} (1991) 3111.

\bibitem{ddv92}C. Destri and H.J. de Vega, 
``New thermodynamic Bethe ansatz equations without strings,'' 
{\emph Phys.Rev.Lett.}{\bf 69 }(1992) 2313-2317.
[hep-th/9203064].



\bibitem{ddv95}C. Destri and H.J. de Vega, 
``Unified approach to thermodynamic Bethe Ansatz and finite size corrections for lattice models and field theories,'' 
\emph{Nucl. Phys.} \textbf{B438} (1995) 413-454,
[hep-th/9407114].


\bibitem{fioravanti}D. Fioravanti, A. Mariottini, E. Quattrini and F. Ravanini, 
``Excited state Destri-De Vega equation for Sine-Gordon and restricted Sine-Gordon models,''
 \emph{Phys. Lett.} \textbf{B390} (1997) 243-251, 
 [hep-th/9608091].
\bibitem{ddv97}C. Destri and H.J. de Vega, 
``Nonlinear integral equation and excited states scaling functions in the sine-Gordon model,''
\emph{Nucl. Phys.} \textbf{B504} (1997) 621-664,
 [hep-th/9701107].
\bibitem{FRT1} G. Feverati, F. Ravanini and G. Tak\'acs, 
``Nonlinear integral equation and finite volume spectrum of Sine-Gordon theory,''
\emph{Nucl. Phys.} \textbf{B540} (1999) 543-586,
[hep-th/9805117].
\bibitem{FRT2}G. Feverati, F. Ravanini and G. Tak\'acs, 
``Truncated conformal space at c = 1, nonlinear integral equation and quantization rules for multi - soliton states,''
\emph{Phys. Lett.} \textbf{B430} (1998) 264-273,
[hep-th/9803104]
\bibitem{FRT3}G. Feverati, F. Ravanini and G. Tak\'acs, `` 	
Scaling functions in the odd charge sector of sine-Gordon / massive Thirring theory,''
\emph{Phys. Lett.} \textbf{B444} (1998) 442-450,
[hep-th/9807160].
\bibitem{FevPhd} G. Feverati, ``Finite volume spectrum of sine-Gordon model and its restrictions (Phd Thesis),''
[hep-th/0001172].
\bibitem{BJsg}
  Z.~Bajnok and R.~A.~Janik,
  ``Classical limit of diagonal form factors and HHL correlators,''
  JHEP {\bf 1701} (2017) 063
  [arXiv:1607.02830 [hep-th]].
  \bibitem{Boos1}
  H.~Boos,
  ``Fermionic basis in conformal field theory and thermodynamic Bethe ansatz for excited states,''
  SIGMA {\bf 7} (2011) 007
  [arXiv:1010.0858 [hep-th]].
  \bibitem{BoosSmir16}
  H.~Boos and F.~Smirnov,
  J.\ Phys.\ A {\bf 51} (2018) no.37,  374003
  doi:10.1088/1751-8121/aad4bb
  [arXiv:1610.09537 [hep-th]].
  %
  \bibitem{DotFat}
  V.~S.~Dotsenko and V.~A.~Fateev,
  ``Conformal Algebra and Multipoint Correlation Functions in Two-Dimensional Statistical Models,''
  Nucl.\ Phys.\ B {\bf 240} (1984) 312.
  %
  \bibitem{ZamiM}
  A.~B.~Zamolodchikov,
  ``Mass scale in the sine-Gordon model and its reductions,''
  Int.\ J.\ Mod.\ Phys.\ A {\bf 10} (1995) 1125.
  %
  \bibitem{BHhnlie}
  J.~Balog and A.~Hegedus,
  ``$AdS_5\times S^5$ mirror TBA equations from Y-system and discontinuity relations,''
  JHEP {\bf 1108} (2011) 095
  [arXiv:1104.4054 [hep-th]].
  %
  \bibitem{SSh}
  A.~Hegedus,
  ``Finite size effects in the SS model: Two component nonlinear integral equations,''
  Nucl.\ Phys.\ B {\bf 679} (2004) 545
  [hep-th/0310051].
  %
  \bibitem{KLsun}
  V.~Kazakov and S.~Leurent,
  Nucl.\ Phys.\ B {\bf 902} (2016) 354
  [arXiv:1007.1770 [hep-th]].
  %
  \bibitem{GKVsu2}
  N.~Gromov, V.~Kazakov and P.~Vieira,
  ``Finite Volume Spectrum of 2D Field Theories from Hirota Dynamics,''
  JHEP {\bf 0912} (2009) 060
  [arXiv:0812.5091 [hep-th]].

\bibitem{klassme}T. Klassen and E. Melzer, ``Sine-Gordon not equal to massive Thirring, and related heresies,''
\emph{Int. J. Mod. Phys}\textbf{. A8} (1993) 4131-4174,
 [hep-th/9206114].


 
 
 
 


\end{thebibliography}
\end{document}